\documentstyle[seceq,preprint,mbf]{ptptex}

\def\slash#1{\ooalign{\hfil/\hfil\crcr$#1$}}

\def\tr{\mathop{\hbox{tr}}}

\def\XZ{{\mbf Z}}

\def\XQ{{\mbf Q}}
\def\XM{{\mbf M}}

\def\XD{{\mbf D}}
\def\XS{{\mbf S}}

\def\XU{{\mbf U}}
\def\XX{{\mbf X}}
\def\XG{{\mbf G}}

\def\dt{\!\cdot\!}
\def\nn{\nonumber\\}

\def\calL{{\cal L}}
\def\calA{{\cal A}}
\def\calD{{\cal D}}
\def\calN{{\cal N}}
\def\hatD{{\hat\calD}}

\def\slashD{{\hat{\slash\calD}}}

\def\hatR{{\hat{\cal R}}}
\def\hatG{{\hat G}}
\def\calF{{\cal F}}

\def\calM{{\cal M}}
\def\calP{{\cal P}}
\def\calR{{\cal R}}
\def\calY{{\cal Y}}
\def\half{\hbox{\large ${1\over2}$}}
\def\myfrac#1#2{\hbox{\large ${#1\over#2}$}}
\def\T{{\rm T}}

\def\pr#1{{#1''}}

\def\6#1{{\underline{#1}}}
\def\m6#1{{\underline{#1}\,}}
\def\hA{{\cal A}}  \def\hF{{\cal F}}
   
\newdimen\Tdim
\def\ispan{{\setbox0=\hbox{i}%
\Tdim\ht0\advance\Tdim\dp0\rule[-\dp0]{0pt}{\Tdim}}}
\def\jspan{{\setbox0=\hbox{j}%
\Tdim\ht0\advance\Tdim\dp0\rule[-\dp0]{0pt}{\Tdim}}}
\def\Tspan#1{{\setbox0=\hbox{#1}%
\Tdim\ht0\advance\Tdim\dp0\advance\Tdim1.5ex\rule[-\dp0]{0pt}{\Tdim}\box0}}
\def\ttp{\tilde t}
\def\tv{{\tilde v}}
\def\ty{\tilde Y}
\def\tV{\tilde V}
\def\V{{\mbf V}}
\def\L{{\mbf L}}
\def\H{{\mbf H}}
\def\N#1{{#1}^{\rm N}{}}
\def\fprm{'}
\def\sprm{''}
\def\tprm{'''}
\preprintnumber[3cm]{
KUNS-1693\\ hep-th/0010288}

\title{
Off-Shell ${\mbf\mbfplus d=5}$ Supergravity Coupled to \\
a Matter-Yang-Mills System}

\author{
Taichiro {\sc Kugo}\footnote{E-mail:
kugo@gauge.scphys.kyoto-u.ac.jp}
and Keisuke {\sc Ohashi}\,\footnote{E-mail:
keisuke@gauge.scphys.kyoto-u.ac.jp}}

\inst{
Department of Physics, Kyoto University, Kyoto 606-8502, Japan}

\recdate{
November 5, 2000}

\abst{
We present an off-shell formulation of a matter-Yang-Mills system 
coupled to supergravity in five-dimensional space-time. We give an 
invariant action for a general system of vector multiplets and 
hypermultiplets coupled to supergravity as well as the supersymmetry 
transformation rules. All the auxiliary fields are retained, so
that the supersymmetry transformation rules remain unchanged even when 
the action is changed.} 

\begin{document}

\maketitle

\section{Introduction}

It is a revolutionary and interesting idea that our 
four-dimensional world may be a `3-brane' embedded in a 
higher-dimensional spacetime. In order to investigate various problems 
seriously in such brane world scenarios, however, we need to understand
supergravity theory in five dimensions.\cite{ref:sugra5,ref:AFMR} 

We are interested in five-dimensional space-time since it provides us 
with the simplest case in which we have a single extra dimension. 
Also, in a more realistic situation, it is believed that M-theory, 
whose low energy effective 
theory is described by eleven-dimensional supergravity, is 
compactified on Calabi-Yau 3-folds and that it can then be described 
by effective five-dimensional supergravity 
theories.\cite{ref:Witten,ref:Lukas} 

In the framework of the `on-shell formulation' (that is, the formulation in 
which there are no auxiliary fields and hence the supersymmetry 
algebra closes only on-shell), 
G\"unayden, Sierra and Townsend\cite{ref:GST} (GST) proposed such a 
five-dimensional supergravity theory which contains general 
Yang-Mills/Maxwell vector multiplets. Their work was extended 
recently by Ceresole and Dall'Agata\cite{ref:CDA} to a rather general 
system containing also tensor (linear) multiplets and hypermultiplets.  

However, in various problems we need an off-shell formulation containing 
the auxiliary fields with which the supersymmetry transformation laws 
are made system-independent and the algebra closes without 
using equations of motion. For instance, 
Mirabelli and Peskin\cite{ref:MP} were able to give 
a simple algorithm based on an off-shell formulation for finding how to 
couple bulk fields to boundary fields in a work in which they 
considered a five-dimensional super Yang-Mills theory compactified on 
$S^1/Z_2$. They clarified how supersymmetry breaking occurring on one 
boundary is communicated to another. Moreover, if we wish to study
problems by adding D-branes to such a system, then, without an 
off-shell formulation, we must find a new supersymmetry transformation 
rule of the bulk fields each time that we add new branes, since 
the supersymmetry transformation law in the on-shell 
formulation depends on the Lagrangian of the system. 

A 5D supergravity tensor calculus for constructing an off-shell formulation 
has been given by Zucker.\cite{ref:Zucker} 
In a previous paper,\cite{ref:KO1} which we refer to as I henceforth, 
we derived a more complete tensor calculus using dimensional reduction 
from 6D superconformal tensor calculus.\cite{ref:BSVP}
Tensor calculus gives a set of rules in  off-shell supergravity: 
i) transformation laws of the various types of supermultiplets; 
ii) composition laws of multiplets from multiplets; and iii) invariant 
action formulas. In this paper, we construct an action for a general 
system of vector multiplets and hypermultiplets coupled to supergravity 
based on the tensor calculus presented in I. 
 This is, in principle, a
straightforward task (containing no trial-and-error steps). 
Nevertheless it requires  considerable computations to 
simplify the form of the action and transformation laws; 
in particular, we must perform a change of variables in order to make 
the Rarita-Schwinger term canonical by solving the mixing between the 
gravitino and matter fermion fields. 

In \S2, we present an invariant action for the system of vector 
multiplets. Although a certain index must be restricted to be of an 
Abelian group in order for the tensor calculus formulas to be applicable, 
we find that the action can in fact be generalized to non-Abelian cases 
by a slight modification.
The action for the system of hypermultiplets is next given in \S3, where 
the mass term is also included. In \S4, we combine these two systems and 
make a first step to simplifying the form of the total action.
In \S5, we fix the dilatation gauge and perform a change of variables 
to obtain the final form of the action, in which both the Einstein and 
Rarita-Schwinger terms take canonical forms. 
This gauge fixing and change of variables modify the supersymmetry 
transformation into a combination of the original 
supersymmetry and other transformations, which are carried out in \S6. 
In \S7, we give comments on i) the relation to the independent variables 
used in GST, ii) compensator components in the hypermultiplets, iii) the
gauging of $SU(2)_R$ and $U(1)_R$, and iv) the scalar potential term in the 
action. We conclude in \S8.
Appendix A gives a technical proof for the existence of a representation 
matrix. In Appendix B, we explicitly show how the manifold 
$U(2,n)/U(2)\times U(n)$ is obtained as a target space of the hypermultiplet 
scalar fields for the case of two (quaternion) compensators.\cite{ref:BS}

In this paper, we do not give the tensor calculus formulas presented 
in our previous paper I, but we freely refer to the equations given there.
For instance, (I\,2$\cdot$3) denotes Eq.~(2$\cdot$3) in I. \
For clarity, however, we list in Table~\ref{table:1} the field contents 
of the Weyl multiplets, vector multiplets and hypermultiplets, which we 
deal with in this paper. (The dilatation gauge field $b_\mu$ and spin 
connection $\omega_\mu^{ab}$ are also listed, although they are 
dependent fields.) \ The notation is the same as in I, 
with one exception: Here we use $\chi^i$ to denote the auxiliary 
fermion component of the Weyl multiplet denoted by $\tilde 
\chi^i$ in I.
\renewcommand{\arraystretch}{1.3}
\begin{table}[tb]
\caption{Field contents of the multiplets.}
\label{table:1}
\begin{center}
\begin{tabular}{ccccc} \hline \hline
    field      & type   & restrictions & {\it SU}$(2)$ & Weyl-weight
    \\ \hline 
\multicolumn{5}{c}{{\it Weyl multiplet}\phantom{MMM}} \\ \hline
\Tspan{$e_\mu{}^a$} &   boson    & f\"unfbein    & \bf{1}    &  $ -1$     \\  
$\psi^i_\mu$  &  fermion  & {\it SU}$(2)$-Majorana & \bf{2}  &$-\myfrac12$ \\  
\Tspan{$V^{ij}_\mu$}    &  boson    & $SU(2)$ gauge field\ 
$V_\mu^{ij}=V_\mu^{ji}=V^*_{\mu ij}$ 
& \bf{3}&0\\ 
$A_\mu$    &  boson    & gravi-photon $A_\mu=e^z_5e_\mu^5$   & \bf{1}&0 \\ 
$\alpha$    &  boson    & `dilaton' $\alpha=e_5^z$   & \bf{1}& 1 \\ 
$t^{ij}$    &  boson    & $t^{ij}=t^{ji}=t^*_{ij}\ (=-V_5^{ij})$  
 & \bf{3} & 1 \\ 
$v_{ab}$&boson& real tensor\ $v_{ab}=-v_{ba}$&\bf{1}&1 \\
$\chi^i$  &  fermion  & {\it SU}$(2)$-Majorana & \bf{2}    &\myfrac32 \\  
$C$    &  boson    & real scalar & \bf{1} & 2 \\ \hline
\Tspan{$b_\mu$} & boson &  $\XD$ gauge field $b_\mu=\alpha^{-1}\partial_\mu\alpha$ 
& \bf{1} & 0 \\
$\omega_\mu{}^{ab}$ &   boson    & spin connection & \bf{1}    &   0 \\ \hline 
\multicolumn{5}{c}{{\it Vector multiplet}\phantom{MMM}} \\ \hline
$W_\mu$      &  boson    & real gauge field   &  \bf{1}    &   0     \\
$M$& boson & real scalar,\quad $M=-W_5$& \bf{1} & 1 \\ 
$\Omega^i$      &  fermion  &{\it SU}$(2)$-Majorana  & \bf{2} &\myfrac32 \\  
$Y_{ij}$    &  boson    & $Y^{ij}=Y^{ji}=Y^*_{ij}$   & \bf{3} & 2 \\ \hline
\multicolumn{5}{c}{{\it Hypermultiplet}\phantom{MMM}} \\ \hline
\Tspan{$\hA_i^\alpha$}     &  boson & 
$\hA^i_\alpha=\varepsilon^{ij}\hA_j^\beta\rho_{\beta\alpha}=-(\hA_i^\alpha)^*$ &\bf{2}& \myfrac32  \\  
$\zeta^\alpha$    &  fermion  & $\bar\zeta^\alpha\equiv(\zeta_\alpha)^\dagger\gamma_0 = \zeta^{\alpha\T}C$ 
& \bf{1}  & 2 \\ 
\rule[-2mm]{0pt}{5mm} {$\hF_i^\alpha$}  &  boson    & 
$ \hF^i_\alpha=-(\hF_i^\alpha)^*$  &  \bf{2}   & \myfrac52 \\ \hline
\end{tabular}
\end{center}
\end{table}

\section{Vector multiplet action}

%

Let $\V^I\equiv(M^I,\,W_\mu^I,\,\Omega^{I\,i},\,Y^{I\,ij})$ $(I=1,2,\cdots,n)$ be
vector multiplets of a gauge group $G$, which we assume to be 
given generally by a direct product of simple groups 
$G_i$ and $U(1)$ groups:
\begin{equation}
G=\prod_iG_i\times\prod_xU(1)_x. \qquad (G_i:\ \hbox{simple})
\end{equation}
The structure constant $f_{IJ}{}^K$ of $G$, \ 
$[t_I,\,t_J]=-f_{IJ}{}^Kt_K$,
is nonvanishing only when $I,J$ and $K$ all 
belong to a common simple factor group $G_i$, and then it is the same 
as the structure constant of the simple group $G_i$.  
The gauge coupling constants can, of course, be different for each factor 
group $G_i$ and $U(1)_x$, but for simplicity, we write
the $G$ transformation of $M^I$, for instance, in the form
$\delta_G(\theta)M^I=g[\theta,M]^I=-gf_{JK}{}^I\theta^JM^K$. 
The quantity $g$ here, therefore,
should be understood as representing the coupling constant $g_i$ of 
$G_i$ when $I\in G_i$. ($M$ here in $[\theta,M]$ denotes 
a matrix such that $M=M^It_I$.)

In addition to these $\V^I\ (I=1,2,\cdots,n)$, we have a 
special vector multiplet called the `central charge vector multiplet', 
which consists of the dilaton $\alpha=e_5^z$ and the gravi-photon 
$A_\mu=e_5^ze_\mu^5$ among the Weyl multiplet:
\begin{equation}
\V^{I=0}=(M^{I=0},\,W^{I=0}_\mu,\,\Omega^{I=0\,i},\,Y^{I=0\,ij})
\equiv(\alpha,\,A_\mu,\,0,\,0).
\label{eq:I=0vector}
\end{equation}
We henceforth extend the group index $I$ to run from 0 to $n$ and use 
$I=0$ to denote this central charge vector multiplet as written here. 
Corresponding to this extension, the gauge group $G$ should also be 
understood to include the central charge $\XZ$ as one of the 
Abelian $U(1)_x$ factor groups. 
Note that the fermion and auxiliary field 
components of this multiplet are zero: $\Omega^{I=0}=Y^{I=0}=0$. Thus the 
number of  scalar and vector components is each $n+1$, while the number 
of $\Omega$ and $Y$ components is each $n$, at this stage. (Below the 
number of scalar components is reduced by 1 through $\XD$-gauge fixing.)

In I, we show that we can construct a linear multiplet 
$\L=(L^{ij},\,\varphi^i,\,E_a,\,N)$, denoted by $f(\V)$, from vector 
multiplets $\V^I$ using any {\it homogeneous quadratic} polynomial in $M^I$,
\begin{equation}
f(M)= \half f_{IJ} M^IM^J,  
\end{equation}
where $I,J$ run from 0. The vector component $E_a$ of a linear multiplet is 
subject to a `divergenceless' constraint, and it can be replaced by an 
unconstrained anti-symmetric tensor (density) field $E^{\mu\nu}$ when $\L$ 
is completely 
neutral under $G$. The explicit expression for the components of 
this multiplet, $\L=f(\V)$, $L^{ij},\,\varphi^i,\,E_a,\,N$ and 
$E^{\mu\nu}$, in terms of those of $\V^I$ is given in 
Eqs.~(I\,5$\cdot$3) and (I\,5$\cdot$5).
We also have the V-L action formula in Eq.~(I\,5$\cdot$7), which gives an 
invariant action for any pair consisting of an Abelian vector multiplet 
$\V=(M,W_\mu,\Omega^{i},Y^{ij})$ and a 
linear multiplet 
$\L=(L^{ij},\,\varphi^i,\,E_a\, (\hbox{or}\, E^{\mu\nu}),\,N)$:
\begin{eqnarray}
e^{-1}{\cal L}_{\rm VL}
&=&Y^{ij}L_{ij}+2i\bar\Omega\varphi+2i\bar\psi^i_a\gamma^a\Omega^jL_{ij} 
+\myfrac12M(N-2i\bar\psi_b\gamma^b\varphi
  -2i\bar\psi^i_a\gamma^{ab}\psi^j_b L_{ij}) \nn
&&{}
-\myfrac12W_a(E^a-2i\bar\psi_b\gamma^{ba}\varphi
  +2i\bar\psi^i_b\gamma^{abc}\psi^j_cL_{ij}) \,.
\label{eq:Act.LV}
\end{eqnarray}
This formula is valid only when the liner multiplet $\L$ 
carries no gauge group charges or is charged 
only under the abelian group of this vector multiplet $\V$. 
When the linear multiplet carries no charges, the constrained 
component $E^a$ can be replaced by the unconstrained anti-symmetric 
tensor $E^{\mu\nu}$, and the action formula (\ref{eq:Act.LV}) can be 
rewritten in a simpler form:
\begin{eqnarray}
e^{-1}{\cal L}_{\rm VL}
&=&Y^{ij}L_{ij}+2i\bar\Omega\varphi+2i\bar\psi^i_a\gamma^a\Omega^jL_{ij} 
+\myfrac12M(N-2i\bar\psi_b\gamma^b\varphi
  -2i\bar\psi^i_a\gamma^{ab}\psi^j_b L_{ij}) \nn
&&{}
+\myfrac14e^{-1}F_{\mu\nu}(W)E^{\mu\nu}.
\label{eq:Act.E}
\end{eqnarray}

\def\I{A}
Now we use this invariant action formula (\ref{eq:Act.E}) to construct a
general action for our set of vector multiplets $\{ \V^I \}$. 
Since this formula applies only to an Abelian vector multiplet $\V$, we first choose all 
the {\it Abelian} vector multiplets $\{ \V^\I \}$ from $\{ \V^I \}$, 
and, for each abelian index $\I$ we prepare a $G$-invariant 
quadratic polynomial $f_\I(M)$ to construct a neutral linear multiplet 
$\L_\I=f_\I(\V)$ using Eqs.~(I\,5$\cdot$3) and (I\,5$\cdot$5).
We apply the V-L action formula (\ref{eq:Act.E}) to these pairs of 
$\V^\I$ and $\L_\I=f_\I(\V)$ and sum over all the Abelian indices $A$. 
Then we rewrite super-covariantized quantities like $\hat G_{ab}(W)$, 
$\hat\calD_aM^I$, $\hat\calD_a\Omega^I$, etc., as non-supercovariantized 
quantities:
\begin{eqnarray}
&&\hatG^I_{ab}(W)=G^I_{ab}(W)+4i\bar\psi_{[a}\gamma_{b]}\Omega^I, \nn
&&\hatD_a\Omega^{Ii}
=\calD_a\Omega^{Ii}
+(\myfrac14\gamma\dt G^I(W) + \half{\slash{\calD}}M^I-Y^I)\psi^i_a
+i\gamma^{bc}\psi^i_a(\bar\psi_b\gamma_c\Omega^I)-i\gamma^b\psi^i_a(\bar\psi_b\Omega^I), \nn
&&\hatD_aM^I=\calD_aM^I-2i\bar\psi_{a}\Omega^I. 
\end{eqnarray}
Here, $\calD_\mu$ is the usual covariant derivative, which is covariant 
only with respect to the homogeneous transformations 
$\XM_{ab},\,\XU_{ij},\,\XD$ and $\XG$. Then, interestingly, many 
cancellations occur, and the resultant expression is no more complicated 
than that written with supercovariantized quantities. Using the notation
\begin{equation}
f_\I\equiv f_\I(M)=\half f_{\I,JK}M^J M^K, \quad 
f_{\I,J} \equiv{\partial f_\I\over\partial M^J }\,,
\quad f_{\I,JK}\equiv{\partial^2f_\I\over\partial M^J\partial M^K }\,,
\end{equation}
the result is given by
\begin{eqnarray}
&&\hspace{-2em}Y^{\I ij}L_{\I ij}+2i\bar\Omega^\I\varphi_\I 
+2i\bar\psi^i_a\gamma^a\Omega^{\I j}L_{\I ij}\nn
&=&f_\I\left(
2Y^\I\dt t -4i\bar\psi\dt\gamma t \Omega^\I -8i\bar\Omega^\I\chi\right) \nn
&&{}+f_{\I,J}
\left(\begin{array}{c}
-Y^\I\dt Y^J +
2i\bar\Omega^\I({\slash{\calD}}-\half\gamma\dt v+t)\Omega^J 
+ i\bar\Omega^\I\gamma^a(\half\gamma\dt G+\slash{\calD}M)^J \psi_a \\
{}-2(\bar\Omega^{\I}\gamma^a\gamma^{bc}\psi_a)(\bar\psi_b\gamma_c\Omega^{J})
+2(\bar\Omega^{\I}\gamma^a\gamma^{b}\psi_a)(\bar\psi_b\Omega^{J})
\end{array} \right) \nn
&&{}+f_{\I,JK}
\left(\begin{array}{c}
-2i\bar\Omega^\I(\myfrac14\gamma\dt G-\half{\slash{\calD}}M+Y)^J\Omega^K
-i\bar\Omega^JY^\I\Omega^K \\
{}+2(\bar\Omega^\I\gamma^{ab}\Omega^{J})(\bar\psi_a\gamma_b\Omega^{K}) 
+2(\bar\Omega^\I\gamma^{a}\Omega^{J})(\bar\psi_a\Omega^{K}) \\
{}+2(\bar\psi^i\dt\gamma\Omega^{\I j})(\bar\Omega_{(i}^J\Omega_{j)}^K) 
\end{array} \right),
\label{eq:YL}
\end{eqnarray}
\begin{eqnarray}
&&\hspace{-2em}\myfrac12M^\I(N_\I-2i\bar\psi_b\gamma^b\varphi_\I
  -2i\bar\psi^i_a\gamma^{ab}\psi^j_b L_{\I ij})
+\myfrac14e^{-1}F^\I_{\mu\nu}(W)E_\I^{\mu\nu} \nn
&=&\half f_{\I,J}\calD^aM^\I(\calD_aM^J-2i\bar\psi_b\gamma^a\gamma^b\Omega^J)\nn
&&{}+\half f_\I M^\I
\left(\begin{array}{c}
-4C-16t\dt t
-\myfrac1{2\alpha}F_{ab}(A)(4v^{ab}+i\bar\psi_c\gamma^{abcd}\psi_d) \\
{}+8i\bar\psi\dt\gamma\chi 
 +4i\bar\psi_a\gamma^{ab}t\psi_b 
\end{array} \right)\nn
&&{}+\half f_{\I,J}M^\I
\left(\begin{array}{c}
 4t\dt Y^J-16i\bar\Omega^J\chi-8i\bar\psi\dt\gamma t\Omega^{J}
+2ig[\bar\Omega,\,\Omega]^J \\
{}-\half G^{Jab}(W)(4v^{ab}+i\bar\psi_c\gamma^{abcd}\psi_d)
\end{array} \right)\nn
&&{}+\half f_{\I,JK}M^\I\left(\begin{array}{c}
 -\myfrac14G^J(W)\dt G^K(W)+\myfrac12\calD_aM^J\calD^aM^K
 -Y^J\dt Y^K \\
+2i\bar\Omega^J(\slash{\calD}-\half\gamma\dt v+t )\Omega^K 
+i\bar\psi_a(\gamma\dt G-2\slash{\calD}M)^J\gamma^a\Omega^K \\
 +(\bar\Omega^J\gamma^{ab}\Omega^K)(\bar\psi_a\psi_b)
  -2(\bar\psi_a^i\gamma^{ab}\psi_b^j)(\bar\Omega^J_i\Omega^K_j) \\
 -4(\bar\psi_a\gamma^{abc}\Omega^K)(\bar\psi_b\gamma_c\Omega^K) 
 +4(\bar\psi_{[a}\gamma_{b]}\Omega^J)(\bar\psi^a\gamma^b\Omega^K) \\
  -2(\bar\psi_a\Omega^J)(\bar\psi^a\Omega^K)
\end{array}\right) \nn[-1ex]
&&-\myfrac14G^\I_{\mu\nu}(W)\bigl(f_\I(4v^{\mu\nu}+i\bar\psi_\rho\gamma^{\mu\nu\rho\sigma}\psi_\sigma)
+if_{\I,JK}\bar\Omega^J\gamma^{\mu\nu}\Omega^K \nn
&&\qquad \qquad \qquad\qquad 
{}+f_{\I,J}(G^{J\mu\nu}(W)-2i\bar\psi_\lambda\gamma^{\mu\nu}\gamma^\lambda\Omega^J)\bigr)\nn
&&{}-e^{-1}\myfrac14f_{\I,JK}\epsilon^{\lambda\mu\nu\rho\sigma}W_{\lambda}^\I
F^J_{\mu\nu}(W)F^K_{\rho\sigma}(W)\ .
\label{eq:MN}
\end{eqnarray}
Here and throughout this paper, we use the following convention for the 
$SU(2)$ triplet quantities $X^{ij}$, like $t^{ij}$, $Y^{Iij}$ and $V_\mu 
^{ij}$: If their $SU(2)$ indices are suppressed, they 
represent the matrix $X^i{}_j$, so that $X\psi^i$, when acting on an 
$SU(2)$ spinor $\psi^i$ like $\Omega^{Ii}$, represents $X^i{}_j\psi^j$,
 and, similarly,  $X\psi_i=X_{ij}\psi^j$, as obtained by lowering 
the index $i$ on both sides. $X\dt\,Y$, for two triplets $X$ and $Y$, 
represents 
$\tr(XY)=X^i{}_jY^j{}_i=-X^{ij}Y_{ji}=-X^{ij}Y_{ij}$, and $X\dt X$ is 
also written $X^2$. For instance, $\bar\Omega^AY^J\Omega^K$ in the above
represents $\bar\Omega^{Ai}Y^J\Omega^K_i=\bar\Omega^{Ai}Y^J_{ij}\Omega^{Kj}$.

The action is given by the sum of Eqs.~(\ref{eq:YL}) and (\ref{eq:MN}), 
where the indices $J$ and $K$ run over the whole group $G$, while the 
(external) index $\I$ of $f_\I(M)$ is restricted to run only over the 
Abelian subset of $G$. However, interestingly, this action can be shown 
to be {\it totally symmetric} with respect to the three indices $\I,J$ 
and $K$ of $f_{\I,JK}$ if $J$ and $K$ are also restricted to the Abelian 
indices. In view also of the fact that this action formula itself gives 
an invariant action, including the case of non-Abelian indices for $J$ 
and $K$, we suspect that this action gives an invariant action even if 
we extend the the index $\I$ of $f_\I(M)$ to $I$ running over the whole 
group $G$. In that case, the function $f_I(M)$ for the indices $I$ 
belonging to the non-Abelian factor groups $G_i$ of $G$ should, of course,
be a function giving the adjoint representation of $G_i$ to satisfy the 
$G$ invariance, and the Chern-Simons term should also be generalized to 
the corresponding one. 
(A similar situation also exists in the 6D 
case.\cite{ref:BSVP}) \ Then the product $M^If_I(M)$ becomes a general 
$G$-invariant {\it homogeneous cubic} polynomial in $M$, which, with 
a sign change, is called a 
`norm function' and denoted $\calN(M)$, following G\"unayden, Sierra 
and Townsend:\cite{ref:GST} 
\begin{equation}
\calN(M)\equiv c_{IJK}M^IM^JM^K \ (= -M^If_I(M)). 
\end{equation}
Here the coefficient $c_{IJK}$ is totally symmetric with respect to the 
indices. Now the resultant action is characterized solely by this
cubic polynomial $\calN(M)$, and we find the vector multiplet action
\begin{eqnarray}
&&\hspace{-2em}e^{-1}\calL_{\rm VL} \nn
&=&
+\half \calN\left(4C+16t\dt t
+\myfrac1{2\alpha}F_{ab}(A)(4v^{ab}+i\bar\psi_c\gamma^{abcd}\psi_d)
-8i\bar\psi\dt\gamma\chi 
 -4i\bar\psi_a\gamma^{ab}t\psi_b \right)\nn
&&{}-\calN_I \left(\begin{array}{c}
 2t\dt Y^I-8i\bar\Omega^I\chi-4i\bar\psi\dt\gamma t\Omega^{I}
+ig[\bar\Omega,\,\Omega]^I \\
{}- G_{ab}^I(W)(v^{ab}+\myfrac{i}4\bar\psi_c\gamma^{abcd}\psi_d)
\end{array} \right)\nn
&&{}-\half \calN_{IJ}\left(\begin{array}{c}
 -\myfrac14G^I(W)\dt G^J(W)+\myfrac12\calD_aM^I\calD^aM^J
 -Y^I\dt Y^J \\
+2i\bar\Omega^I(\slash{\calD}-\half\gamma\dt v+t )\Omega^J 
+i\bar\psi_a(\gamma\dt G(W)-2\slash{\calD}M)^I\gamma^a\Omega^J \\
-2(\bar\Omega^{I}\gamma^a\gamma^{bc}\psi_a)(\bar\psi_b\gamma_c\Omega^{J})
+2(\bar\Omega^{I}\gamma^a\gamma^{b}\psi_a)(\bar\psi_b\Omega^{J})
\end{array}\right) \nn
&&{}-\calN_{IJK}\left(\begin{array}{c}
-i\bar\Omega^I(\myfrac14\gamma\dt G(W)+Y)^J\Omega^K
 \\ {}+\myfrac23(\bar\Omega^I\gamma^{ab}\Omega^{J})(\bar\psi_a\gamma_b\Omega^{K}) 
+\myfrac23(\bar\psi^i\dt\gamma\Omega^{Ij})(\bar\Omega_{(i}^J\Omega_{j)}^K) 
\end{array}\right) \nn
&&{}+e^{-1}{\cal L}_{\rm C\hbox{-}S}\ ,
\label{eq:YM}
\end{eqnarray}
where $\calN_I=\partial\calN/\partial M^I$, 
$\calN_{IJ}=\partial^2\calN/\partial M^I\partial M^J$, etc., and 
${\cal L}_{\rm C\hbox{-}S}$ is the Chern-Simons term:
\begin{eqnarray}
{\cal L}_{\rm C\hbox{-}S}&=&
\myfrac{1}8 c_{IJK}\epsilon^{\lambda\mu\nu\rho\sigma}W_{\lambda}^I
\bigl(F_{\mu\nu}^J(W)F_{\rho\sigma}^K(W) 
+\myfrac{1}2g[W_\mu,W_\nu]^JF_{\rho\sigma}^K(W) \nn
&&\hspace{8em}{}+\myfrac{1}{10}g^2[W_\mu,W_\nu]^J[W_\rho,W_\sigma]^K\bigr)\,. 
\label{eq:ChernSimons}
\end{eqnarray}

We have checked the supersymmetry invariance of this action for general 
non-Abelian cases as follows. When the gauge coupling $g$ is 
set equal to zero, the action reduces to one with the same form 
as that for the Abelian case, 
and thus the invariance is guaranteed by the above 
derivation. When $g$ is switched on, the covariant derivative $\calD_\mu$ 
comes to include the $G$-covariantization term $-g\delta_G(W_\mu)$, and the 
field strength $F_{\mu\nu}(W)$ comes to include the non-Abelian term 
$-g[W_\mu,W_\nu]$. We, however, can use the variables $\calD_\mu\phi$ ($\phi 
=M^I,\,\Omega^I$) and $F_{\mu\nu}(W)$ as they stand in the action and in the 
supersymmetry transformation laws, keeping these $g$-dependent terms 
implicit inside of them. Then, we have only to keep track of {\it 
explicitly} 
$g$-dependent terms and make sure that these terms vanish in the 
supersymmetry transformation of the action. The explicitly 
$g$-dependent terms in the action are only the term 
$-ig\calN_I[\bar\Omega,\Omega]^I$, aside from those in the 
Chern-Simons term. The Chern-Simons term is special because it 
contains the gauge field $W_\mu^I$ explicitly, and its supersymmetry 
transformation as a whole yields no explicit $g$-dependent terms, as we 
show below. In the supersymmetry transformations $\delta\phi$, explicitly 
$g$-dependent terms do not appear for $\phi=M^I$, $\Omega^I$, 
$G_{\mu\nu}^I(W)$ or $F_{\mu\nu}^I(W)$, but appear only in $\delta 
Y^{Iij}$, $\delta(\calD_\mu M^I)$ and $\delta(\calD_\mu\Omega^I)$. (For 
the latter two, the supersymmetry transformation of $W_\mu$ contained 
implicitly in $\calD_\mu$ produces additional explicitly $g$-dependent 
terms). It is easy to see that all these $g$-dependent terms  cancel 
out in the transformation of the action.

In carrying out such computations,  
it is convenient to use a matrix notation to represent the norm function 
$\calN$. One can show
that, for any $G$-invariant 
$\calN(M)=c_{IJK}M^IM^JM^K$, 
there is a set of hermitian matrices $\{\,T_I\,\}$
which satisfies 
\begin{equation}
c_{IJK}= \myfrac16\tr(T_I\,\{T_J,T_K\})
\label{eq:CIJK}
\end{equation}
and gives a representation of $G$ up to normalization constants $c_i$ 
for each simple factor group $G_i$; that is, 
the rescaled matrices 
$t_I\equiv iT_I/c_{[I]}$, where $c_{[I]}=c_i$ for 
$I\in G_i$ and $c_{[I]}=1$ for $I\in U(1)_x$, 
satisfy 
\begin{equation}
[t_I,\,t_J]= -f_{IJ}{}^K t_K. 
\end{equation}
In Appendix A, we give a simple example of the 
representation of $G$ which realizes these properties. Using the matrix 
notation $\tilde X\equiv X^IT_I$, we have
\begin{eqnarray}
&&\calN\equiv c_{IJK}M^IM^JM^K = \myfrac13\tr(\tilde M^3), \nn
&&\calN_IX^I= \tr(\tilde X\tilde M^2), \qquad 
\calN_{IJ}X^IY^J= \tr(\tilde X\{\tilde Y,\tilde M\}), \nn
&&\calN_{IJK}X^IY^JZ^K= \tr(\tilde X \{\tilde Y,\tilde Z \} ).
\end{eqnarray}
With these expressions, we can simply use cyclic identities for the trace 
instead of referring to various cumbersome identities for $c_{IJK}$ 
resulting from its $G$-invariance property. 
Note the difference from the ordinary matrix notation 
$X\equiv X^It_I$: In the present case we have the relations 
$\widetilde{[X,Y]}=[X,Y]^IT_I=
-f_{IJ}{}^KX^IY^JT_K=[\tilde X,Y]=[X,\tilde Y]$, since 
$f_{IJ}{}^K$ is nonvanishing only when $I,J,K$ belong to a common 
simple factor group $G_i$.

Using this matrix notation for the gauge field $W_\mu^I$ and 
the field strength $F_{\mu\nu}^I$, 
we can define the matrix-valued 1-form as 
$\tilde W\equiv\tilde W_\mu dx^\mu$ and the 2-form as
$\tilde F\equiv\half\tilde F_{\mu\nu}dx^\mu dx^\nu
=d\tilde W-g\widetilde{W^2}$
(where $g\widetilde{W^2}=g\{\tilde W,W\}/2$), 
with which the Chern-Simons 
term (\ref{eq:ChernSimons}) can be rewritten in the form
\begin{equation}
\int{\cal L}_{\rm C\hbox{-}S}\,d^5x = \int\myfrac16\tr\left(
\tilde W\tilde F\tilde F+\myfrac14\{\tilde W,\,g\widetilde{W^2}\}\tilde F
+\myfrac1{10}\tilde W\,g\widetilde{W^2}\,g\widetilde{W^2}
\right).
\end{equation}
For an arbitrary variation of $W_\mu^I$, i.e., $\delta\tilde W=\tilde X$ in 
the matrix-valued 1-form notation, we find $\delta\tilde F=d\tilde X
-g\widetilde{\{W, X\}}$. Using the Bianchi identity $D\tilde F=d\tilde F
-g\widetilde{[W,F]}=0$ and the properties 
$g\widetilde{\{W,X\}}=\{g\tilde W, X\}=\{gW, \tilde X\}$, 
 $g\widetilde{[W,F]}=[g\tilde W, F]=[gW, \tilde F]$ 
and $[g\widetilde{W^2}, W]=[gW^2, \tilde W]=g\widetilde{[W^2,W]}=0$, 
we can show
\begin{eqnarray}
&&\delta\tr\left(
\tilde W\tilde F\tilde F
\right) = \tr\left(3\tilde F\tilde F\tilde X 
- \{\tilde F,\,g\widetilde{W^2}\}\tilde X\right),\nn
&&\delta\tr\left(
\{\tilde W,\,g\widetilde{W^2}\}\tilde F
\right) = \tr\left(4\{\tilde F,\,g\widetilde{W^2}\}\tilde X 
- 2g\widetilde{W^2}\,g\widetilde{W^2}\tilde X\right), \nn
&&\delta\tr\left(
\tilde W\,g\widetilde{W^2}\,g\widetilde{W^2}
\right)= \tr\left( 5g\widetilde{W^2}\,g\widetilde{W^2}\tilde X\right),
\end{eqnarray}
so that the variation of the Chern-Simons term indeed gives no explicitly 
$g$-dependent term, as claimed above:
\begin{equation}
\delta\int{\cal L}_{\rm C\hbox{-}S}\,d^5x =
\int\half\tr(\tilde F\tilde F \delta\tilde W)\,.
\end{equation}

\vspace*{1mm}

\section{Hypermultiplet action}

Now let $\H^\alpha=(\calA^\alpha_i,\,\zeta^\alpha,\,\calF^\alpha_i)$ 
$(\alpha=1,2,\cdots,2r)$ be a set of hypermultiplets 
which belongs to a representation $\rho$ of the gauge group $G$. Under the 
$G$ transformation it transforms as 
$\delta_G(\theta)\H^\alpha=\sum_{I=1}^ng\theta^I\rho(t_I)^\alpha{}_\beta 
\H^\beta$. The ordinary matrix notation used for the vector multiplet in 
the preceding section was, for instance, $M=M^It_I$, and the matrix $t_I$ 
denoted an adjoint representation $\hbox{ad}(t_I)$ of $G$. The 
representation $\rho$ here can, of course, be different from the adjoint 
representation $\hbox{ad}$. However, to avoid cumbersome expressions, we 
simplify the matrix notation and write, e.g., 
$M\calA^\alpha_i=M^\alpha{}_\beta\calA^\beta_i$ to represent 
$\rho(M)^\alpha{}_\beta\calA^\beta_i=M^I\rho(t_I)^\alpha{}_\beta 
\calA^\beta_i$. (Note $M\calA_{\alpha i}=M_{\alpha\beta}\calA^\beta_i$.)

The invariant action for the hypermultiplets is derived in I from the 
action in 6D and is given by Eq.~(I\,4$\cdot$11).\footnote{This 
action can also be derived if we make a 
linear multiplet $\L=d_{\alpha\beta}\H^\alpha\times{\mbf Z}\H^\beta$ from 
the hypermultiplets $\H^\alpha$ and their central-charge transforms 
${\mbf Z}\H^\beta$ by using the formula (I\,5$\cdot$6), and then apply 
the linear multiplet action formula (I\,5$\cdot$9) to it.} Again we 
rewrite the supercovariant derivative $\hatD_\mu$ in terms of the usual 
covariant derivative $\calD_\mu$, which is covariant only with respect 
$\XM_{ab},\,\XU_{ij},\,\XD$ and $\XG$. (Note that covariantization with 
respect to the central charge $\XZ$ transformation is also taken out.) 
Then we obtain the following action for the kinetic term of the 
hypermultiplets:
\begin{eqnarray}
e^{-1}{\cal L}_{\rm kin}&=&\calD^a\calA_i^{\bar \alpha}\calD_a\calA^i_\alpha 
-2i\bar\zeta^{\bar \alpha}\slash{\calD}\zeta_\alpha 
+\myfrac{i}{2\alpha}\bar\zeta^{\bar \alpha}\gamma\dt F(A)\zeta_\alpha 
-i\bar\zeta^{\bar \alpha}\gamma\dt v\zeta_\alpha\nn
&&{}+2ig\bar\zeta^{\bar \alpha}M_\alpha{}^\beta\zeta_\beta 
+\calA^{\bar \alpha}_i(t+gM)^2\calA^i_\alpha 
-4i\bar\psi^i_a\zeta_\alpha\gamma^b\gamma^a\calD^a\calA^{\bar \alpha}_i \nn
&&{}+\left(\begin{array}{c}
   2i\bar\zeta_\alpha\gamma^{ab}{\cal R}_{ab}{}^i(Q)
  -8i\bar\zeta_\alpha\chi^i\\
  +\myfrac{i}{\alpha}\bar\psi^i_a\gamma^{abc}\zeta_\alpha\hat F_{bc}(A)
   -4i\bar\psi^{ia}\gamma^b\zeta_\alpha v_{ab}
  +4i\bar\psi_{aj}\gamma^a\zeta_\alpha t^{ij}\\
  -8ig\bar\Omega^i_\alpha{}^\beta\zeta_\beta+4ig\bar\psi_a^i\gamma^aM_\alpha{}^\beta\zeta_\beta 
 \end{array} \right)\calA^{\bar\alpha}_i\nn
&&{}-2i\bar\psi^{(i}_a\gamma^{abc}\psi_c^{j)}\calA^{\bar\alpha}_j\calD_b\calA_{\alpha i}\nn
&&{}+\left(\begin{array}{c}
  C+\myfrac14{\cal R}(M)+\myfrac{i}2\bar\psi_a\gamma^{abc}{\cal R}_{bc}(Q)\\
  -2i\bar\psi_a\gamma^a\chi 
 +\myfrac1{8\alpha^2}\hat F(A)^2-v^2+2t\dt t\\
  -\myfrac{i}{4\alpha}\bar\psi_a\gamma^{abcd}\psi_b\hat F_{cd}(A)
  +i\bar\psi_a\psi_bv^{ab}-i\bar\psi_a^i\gamma^{ab}\psi^j_bt_{ij}\\
\end{array} \right)\calA^2\nn
&&{}+2gY^{ij}_{\alpha\beta}\calA^{\bar\alpha}_i\calA^\beta_j
+4ig\bar\psi_a^{(i}\gamma^a\Omega^{j)}_{\alpha\beta}\calA^{\bar\alpha}_i\calA^\beta_j\nn
&&{}+2ig\bar\psi^{(i}_a\gamma^{ab}\psi^{j)}_b\calA^{\bar\alpha}_iM_\alpha{}^\beta\calA_{\beta j}
+(1-A^aA_a/\alpha^2)\calF^{\bar\alpha}_i\calF^i_\alpha\nn
&&{}+\bar\psi_a\gamma_b\psi_c\bar\zeta^{\bar\alpha}\gamma^{abc}\zeta_\alpha 
-\myfrac12\bar\psi^a\gamma^{bc}\psi_a\bar\zeta^{\bar\alpha}\gamma_{bc}\zeta_\alpha\ ,
\label{eq:hyperK}
\end{eqnarray}
where the contraction between a pair with a barred index $\bar\alpha$ and 
$\alpha$ is defined as
\begin{equation}
\calA^{\bar\alpha}_i\calA_{\alpha j}\equiv\calA^\beta_id_\beta{}^\alpha\calA_{\alpha j}, 
\qquad \calA^2\equiv 
\calA^{\bar\alpha}_i\calA_\alpha^i,
\qquad \bar\zeta^{\bar\alpha}\zeta_\alpha\equiv \bar\zeta^{\beta}d_\beta{}^\alpha\zeta_\alpha,
\end{equation}
by using the $G$-invariant metric $d_\alpha{}^\beta$ introduced in 
Eqs.~(I\,B$\cdot$22) and (I\,B$\cdot$23). This metric $d_\alpha{}^\beta$ 
is, in its standard form, diagonal and takes the values 
$\pm1$.\cite{ref:dWLVP} Note in the above that 
$(t+gM)^2\calA^i_\alpha=t^i{}_kt^k{}_j\calA^j_\alpha 
+2gM_{\alpha\beta}t^i{}_j\calA^{\beta 
j}+gM_{\alpha\gamma}gM^\gamma{}_\beta\calA^{\beta i}$ with our 
present convention. The hypermultiplets can have masses, and the 
invariant action for the mass term is given by Eq.~(I\,4$\cdot$14),
 which reads 
\begin{eqnarray}
\hspace*{-5mm}
e^{-1}{\cal L}_{\rm mass}&=&
m\eta^{\alpha\beta}\left(\begin{array}{c}
-A^a\calD_a\calA_{\alpha i}\calA^i_\beta-(1-A^aA_a/\alpha^2)\alpha\calF_{\alpha i}\calA_\beta^i\\
-2i\bar\psi_a^i\zeta_\alpha A^a\calA_{\beta i}
+\alpha\calA_{i\alpha}(t+gM)\calA_\beta^i\\
+i(-\alpha\bar\zeta_\alpha\zeta_\beta+A_a\bar\zeta_\alpha\gamma^a\zeta_\beta)\\
+2i\calA_{\alpha i}(-\alpha\bar\psi^i_a\gamma^a\zeta_\beta+\bar\psi^i_a\gamma^{ab}\zeta_\beta A_b)\\
+i\calA_{\alpha i}\calA_{\beta j}(-\alpha\bar\psi^i_a\gamma^{ab}\psi^j_b
+\bar\psi^i_a\gamma^{abc}\psi^j_cA_b)
\end{array}\right).
\label{eq:hypermass}
\end{eqnarray}
(Note that $m$ is a dimensionless parameter, and 
the actual mass is proportional to $m\langle\alpha\rangle$.)
Here $\eta^{\alpha\beta}$ is a symmetric $G$-invariant tensor.\cite{ref:dWLVP}  
Interestingly, this mass term turns out to be automatically included in the 
previous kinetic term action (\ref{eq:hyperK}), and it need not be 
considered separately, provided that we complete the square for the 
terms containing the auxiliary fields $\calF^\alpha_i$ in 
${\cal L}_{\rm kin}+{\cal L}_{\rm mass}$. (Essentially the same observation
is made in Ref.~\citen{ref:dWLVP}.) \ 
Doing so, the 
$\calF^\alpha_i$ terms become
\begin{equation}
(1-A^aA_a/\alpha^2)\tilde\calF^{\bar\alpha}_i\tilde\calF^i_\alpha 
\qquad \hbox{with}\ \ \tilde\calF^{\alpha}_i\equiv\calF^{\alpha}_i+
\half m\alpha(d^{-1})_\gamma{}^\alpha\eta^{\gamma\beta}\calA_{\beta i}\,,
\label{eq:Fsquare}
\end{equation}
and then, all the other terms in ${\cal L}_{\rm mass}$ can be absorbed into 
the kinetic Lagrangian ${\cal L}_{\rm kin}$ if we extend
the gauge index $I$ of the generators $t_I$ acting on the hypermultiplets 
to run also from 0 and introduce 
\begin{equation}
(gt_{I=0})^{\alpha\beta}\equiv\half m (d^{-1})_\gamma{}^\alpha\eta^{\gamma\beta},
\label{eq:undstd0}
\end{equation}
so that $gW_\mu$ in $\calD_\mu$ and $M$ are now understood to be
\begin{eqnarray}
&&gW_\mu
= \sum_{I=1}^n W_\mu^I(gt_I)+A_\mu(gt_0)\,, \nn
&&gM 
= \sum_{I=1}^n M^I(gt_I)+\alpha(gt_0)\,.
\label{eq:undstd}
\end{eqnarray}

\section{First step in rewriting the action}

Now, the invariant action for our Yang-Mills-matter system coupled 
to supergravity is given by the sum ${\cal L}={\cal L}_{\rm VL}[(\ref{eq:YM})]+
{\cal L}_{\rm kin}[(\ref{eq:hyperK})]$, where in ${\cal L}_{\rm kin}$ 
the $\calF^2$ term is replaced by (\ref{eq:Fsquare}), and 
Eq.~(\ref{eq:undstd}) is understood. 

We first note that the auxiliary fields $C$ and $\chi$ appear in the 
action ${\cal L}$ in the form of Lagrange multipliers:
\begin{equation}
C(\calA^2+2\calN)-8i\bar{\chi}(\zeta+\Omega)\,,
\end{equation}
where $\zeta_i$ and $\Omega_i$ are defined as
\begin{equation}
\zeta_i\equiv\calA^{\bar\alpha}_i\zeta_\alpha=\calA^\beta_id_\beta{}^\alpha\zeta_\alpha, 
\qquad \Omega_i\equiv\calN_I\Omega^I_i.
\end{equation}
That is, $\calA^2=-2\calN$ and $\zeta_i=-\Omega_i$ are equations of motion. 
Although we do not use equations of motion, we can rewrite the terms 
multiplied by $\calA^2$, $\calA^2X$, as $-2\calN X$ with the shift 
$C\rightarrow C+X$, and, similarly, we can rewrite the terms $\bar 
X\zeta$ as $-\bar X\Omega$ with the shift $\chi\rightarrow\chi+iX/8$. 
Using this, we replace all the terms containing the factor $\calA^2$ and
all the terms containing the factor 
$\zeta_i=\calA^{\bar\alpha}_i\zeta_\alpha$ in ${\cal L}_{\rm kin}$ by 
those multiplied by $\calN$ and by $\Omega_i$, respectively.

When doing this, we also rewrite the covariant derivative $\calD_\mu$ 
in the following form, separating the terms containing 
gauge fields $b_\mu$ ($=\alpha^{-1}\partial_\mu\alpha$) and $V_\mu^{ij}$:
\begin{eqnarray}
\calD_\mu&=&\nabla_\mu-\delta_D(b_\mu)-\delta_U(V_\mu^{ij})-\delta_M(-2e_\mu{}^{[a}b^{b]}).
\end{eqnarray}
The last term appears because the spin 
connection $\omega_\mu^{ab}$ contains the $b_\mu$ field as 
\begin{eqnarray}
  \omega_\mu^{\ ab}& = &
   \omega_\mu^{0\ ab}+i(2\bar\psi_\mu\gamma^{[a}\psi^{b]} 
    +\bar\psi^a\gamma_\mu\psi^b)-2e_\mu^{\ [a}b^{b]}\,, \nn
   \omega_\mu^{0\ ab} &\equiv& -2e^{\nu[a}\partial_{[\mu}e_{\nu]}^{\ b]}
    +e^{\rho[a}e^{b]\sigma}e_\mu{}^c\partial_\rho e_{\sigma c}\,.
\label{eq:connection}
\end{eqnarray}
Then, the covariant derivative $\nabla_\mu$ is now covariant only with respect
to local-Lorentz and group transformations, and the spin connection 
is that with $b_\mu$ set equal to 0:
\begin{equation}
\nabla_\mu=\partial_\mu-\delta_M(\omega_\mu^{ab}|_{b_\mu=0})-\delta_G(W_\mu).
\end{equation} 
We perform this separation of the $b_\mu$ and $V_\mu^{ij}$ gauge fields
also for $\calR(M)$ and $\calR_{ab}^i(Q)$. This separation also 
yields 
several terms proportional to $\calA^2$ and $\zeta_i$, which also can be
rewritten as terms proportional to $\calN$ and $\Omega_i$ with shifts 
of $C$ and $\chi$.

Thus, we finally define $C'$ and $\chi'$ in terms of 
$C$ and $\chi$ as follows:
\begin{eqnarray}
C\fprm&=&
C+\myfrac14 \calR(M) +\myfrac{i}2\bar\psi_a\gamma^{abc}\calR_{bc}(Q)
-2i\bar\psi_a\gamma^a\chi 
 +\myfrac1{8\alpha^2}\hat F(A)^2 \nn
&&{}-v^2-\myfrac{i}{4\alpha}\bar\psi_a\gamma^{abcd}\psi_b\hat F_{cd}(A)
  +i\bar\psi_a\psi_bv^{ab}-i\bar\psi_a^i\gamma^{ab}\psi^j_bt_{ij} \nn
&&{}+\myfrac94b^2 +\myfrac52 t\dt t +\myfrac32e^{-1}\nabla_\mu(eb^\mu) 
+ \half V^{ij}_aV^a_{ij} 
+i\bar\psi_b^i\gamma^{bac}\psi_c^jV_{a\,ij}\,, \nn
\chi\fprm_i&=&
\chi_i-\myfrac14\gamma^{ab}{\cal R}_{abi}(Q)
  +\myfrac1{8\alpha}\gamma^{abc}\psi_{ai}\hat F_{bc}(A) \nn
&&{} +\half\gamma_b\psi_{ai}v^{ab} +\half t\gamma\dt\psi_i
+ \gamma^a\gamma^b(\half V_b-\myfrac34b_b)\psi_{ai}\,.
\hspace{2em}
\label{eq:Cchiredef}
\end{eqnarray}
We also separate and collect the terms containing $F_{ab}(A)$ and the 
auxiliary fields $v^{ab},\,V_\mu^{ij}$,
\,$t^{ij}$,\,$Y^{Iij},\,\calF^i_\alpha$.
Then the action ${\cal L}$ is found to take the following form at 
this stage:
\begin{eqnarray}
&&\hspace{-5.5em}{\cal L}
={\cal L}'_{\rm hyper}+{\cal L}'_{\rm vector}+{\cal L}_{\rm C\hbox{-}S}+{\cal L}'_{\rm aux}\ , \nn
e^{-1}{\cal L}'_{\rm hyper}&=&\nabla^a\calA_i^{\bar \alpha}\nabla_a\calA^i_\alpha 
-2i\bar\zeta^{\bar \alpha}(\slash{\nabla}+gM)\zeta_\alpha\nn
&&{}
+\calA^{\bar\alpha}_i(gM)^2{}_\alpha{}^\beta\calA_\beta^i
-4i\bar\psi^i_a\gamma^b\gamma^a\zeta_\alpha\nabla_b\calA^{\bar \alpha}_i 
-2i\bar\psi^{(i}_a\gamma^{abc}\psi_c^{j)}\calA^{\bar\alpha}_j\nabla_b\calA_{\alpha i}\nn
&&{}+\calA^{\bar\alpha}_i
\bigl(8ig\bar\Omega^i_{\alpha\beta}\zeta^\beta-4ig\bar\psi_a^i\gamma^aM_{\alpha\beta}\zeta^\beta 
\nn &&{}
\qquad \qquad +4ig\bar\psi_a^{(i}\gamma^a\Omega^{j)}_{\alpha\beta}\calA^\beta_j 
-2ig\bar\psi^{(i}_a\gamma^{ab}\psi^{j)}_bM_{\alpha\beta}\calA^\beta_j\bigr)
\nn
&&{}+\bar\psi_a\gamma_b\psi_c\bar\zeta^{\bar\alpha}\gamma^{abc}\zeta_\alpha 
-\myfrac12\bar\psi^a\gamma^{bc}\psi_a\bar\zeta^{\bar\alpha}\gamma_{bc}\zeta_\alpha\ ,\nn
e^{-1}{\cal L}'_{\rm vector}&=& 
\calN\left(
  -\half{\cal R}(M)|_{b=0}-2i\bar\psi_\mu\gamma^{\mu\nu\rho}\nabla_\nu\psi_\rho 
  +(\bar\psi_a\psi_b)(\bar\psi_c\gamma^{abcd}\psi_d+\bar\psi^a\psi^b)
\right) \nn
&&{}-\calN_I \left(
\begin{array}{c}
   -4i\bar\Omega^I\gamma^{\mu\nu}\nabla_\mu\psi_\nu+2\bar\Omega^I\gamma^{abc}\psi_a\bar\psi_b\psi_c\\{}
   +ig[\bar\Omega,\,\Omega]^I - \myfrac{i}4\bar\psi_c\gamma^{abcd}\psi_dF_{ab}^I(W)
\end{array} \right)\nn
&&{}-\half \left(\calN_{IJ}-{\calN_I\calN_J\over\calN}\right)\left(
 -\myfrac14F^I(W)\dt F^J(W)+\myfrac12\nabla_aM^I\nabla^aM^J\right) \nn
&&{}-\half \calN_{IJ}
\left(\begin{array}{c}
+2i\bar\Omega^I\slash{\nabla}\Omega^J 
+i\bar\psi_a(\gamma\dt F(W)-2\slash{\nabla}M)^I\gamma^a\Omega^J \\
-2(\bar\Omega^{I}\gamma^a\gamma^{bc}\psi_a)(\bar\psi_b\gamma_c\Omega^{J})
+2(\bar\Omega^{I}\gamma^a\gamma^{b}\psi_a)(\bar\psi_b\Omega^{J})
\end{array}\right) \nn
&&{}-\calN_{IJK}\left(\begin{array}{c}
-i\bar\Omega^I\myfrac14\gamma\dt F^J(W)\Omega^K
\\{}+\myfrac23(\bar\Omega^I\gamma^{ab}\Omega^{J})(\bar\psi_a\gamma_b\Omega^{K}) 
+\myfrac23(\bar\psi^i\dt\gamma\Omega^{Ij})(\bar\Omega_{(i}^J\Omega_{j)}^K) 
\end{array}\right) ,
\nonumber
%
%
%
%
\end{eqnarray}
\vspace*{-2ex}
\begin{eqnarray}
e^{-1}{\cal L}'_{\rm aux}&=& 
C\fprm(\calA^2+2\calN) -8i\bar\chi\fprm(\zeta+\Omega) -b^a\nabla_a\calN \nn
&&{}+2\calN\bigl(v-\myfrac1{2\alpha}F(A)+{\calN_I\over4\calN}F^I(W)\bigr)^2 \nn
&&{}-i\bigl(v-\myfrac1{2\alpha}F(A)\bigr)^{\!ab}\!\!
\left(2\calN\bar\psi_a\psi_b
+\bar\zeta^{\bar \alpha}\gamma_{ab}\zeta_\alpha-4\bar\psi_a\gamma_b\Omega 
-\half\calN_{IJ}\bar\Omega^I\gamma_{ab}\Omega^J\right) \nn
&&{}-\calN V_a^{ij}V^a_{ij}
-V_{ij}^a\bigl(2\calA^{\bar\alpha i}\nabla_a\calA_\alpha^j
+4i\bar\Omega^{i}\psi_a^j-i\calN_{IJ}\bar\Omega^{Ii}\gamma_a\Omega^{Jj}\bigr) \nn
&&{}-\half \calN_{IJ}(Y^{ijI}-M^It^{ij})(Y^J_{ij}-M^Jt_{ij}) \nn
&&{}+(Y_{ij}^I-M^It_{ij})(2\calA^i_{\alpha}(gt_I)^{\bar\alpha\beta}\calA_\beta^j
+i\calN_{IJK}\bar\Omega^{Ji} \Omega^{Kj}) \nn
&&{}+\left(1-{A^2/\alpha^2}\right)
(\calF^{\bar\alpha}_i-\calF^{\bar\alpha}_{{\rm sol}\,i})
(\calF^i_\alpha-\calF^{\,i}_{{\rm sol}\,\alpha})\ ,
\label{eq:actionI}
\end{eqnarray}
where $\calF^{\,\alpha}_{{\rm sol}\,i}$ is the solution 
of the equation of motion for $\calF^{\alpha}_i$,
\begin{equation}
\calF^{\,\alpha}_{{\rm sol}\,i}
= -\half \alpha m(d^{-1})_\gamma^\alpha\eta^{\gamma\beta}\calA_{\beta i}
= -\alpha(gt_{I=0})^{\alpha\beta}\calA_{\beta i} =
(gM^0t_0)^\alpha{}_\beta\calA^\beta_i.
\end{equation}
Here it is quite remarkable that all the terms explicitly containing either 
$b_\mu \ (=\alpha^{-1}\partial_\mu\alpha)$ or $F_{\mu\nu}(A)$ have completely 
disappeared from the action, other than ${\cal L}'_{\rm aux}$, except
for the terms contained in the form 
$M^{I}$ and $F^{I}(W)$. That is, $\alpha=M^{I=0}$ and 
$F_{\mu\nu}(A)=F^{I=0}_{\mu\nu}(W)$, which carry the index $I=0$, do not 
appear by themselves, but are only contained in the action 
in a form that is completely symmetric  with the  components
 with $I\geq1$.

\section{Final form of the action}

In view of the action (\ref{eq:actionI}), we note that the Einstein 
term can be made canonical if 
\begin{equation}
\calN(M)=1.
\label{eq:Dgauge}
\end{equation}
$\calN(M)$ is a cubic function of $M^I$, but we fortunately have local 
dilatation $\XD$ symmetry, so that we can take $\calN(M)=1$ as a gauge 
fixing condition for the $\XD$ gauge.\cite{ref:KU2} 

However, the action (\ref{eq:actionI}) is still not in the final form, 
since there remains a mixing kinetic term 
$4i\calN_I\bar\Omega^I\gamma^{\mu\nu}\nabla_\mu\psi_\nu$
between the Rarita-Schwinger field $\psi_\mu^i$ and the gaugino field 
component $\Omega_i=\calN_I\Omega^I_i$. 
If we had superconformal symmetry, 
we could remove the mixing simply by imposing
\begin{equation}
\calN_I\Omega^I_i\equiv\Omega_i=0
\label{eq:Sgauge}
\end{equation}
as a conformal $\XS$ supersymmetry gauge fixing condition. 
Unfortunately, we already fixed the $\XS$ 
gauge when performing the dimensional reduction from 6D to 5D, and thus we 
no longer have such $\XS$ symmetry. Therefore we here must remove the 
mixing by making field redefinitions. The proper Rarita-Schwinger field 
is found to be 
\begin{equation}
\psi^{\rm N}_{\mu i}=\psi_{\mu i}-{1\over3\calN}\gamma_\mu\Omega_i\,.
\label{eq:newRS}
\end{equation}
We also redefine the gaugino fields as
\begin{eqnarray}
\lambda^I_i \equiv\Omega^I_i - {M^I\over3\calN}\Omega_i=
\calP^I_J \Omega^J_i, 
\label{eq:gaugino}
\end{eqnarray}
where $\calP^I_J$ is the projection operator
\begin{equation}
\calP^I_J\equiv\delta^I_J -{M^I\calN_J\over3\calN}\quad \rightarrow\quad 
\calP^I_JM^J=\calP^I_J\calN_I=0\,.
\label{eq:Proj}
\end{equation}
This new gaugino fields $\lambda^I_i$ satisfy
\begin{eqnarray}
\calN_I\lambda^I_i &=& 0\,,
\end{eqnarray}
so that they correspond to the gaugino fields $\Omega_i^I$ which we would 
have had if we could have imposed the $\XS$ gauge fixing condition 
(\ref{eq:Sgauge}). Note, however, that the number of independent 
components of $\lambda^I$ is the same as that of the original $\Omega 
^I$, since the $I=0$ component of the latter vanishes: $\Omega^{I=0}=0$. 
Note also that Eq.~(\ref{eq:gaugino}) and the relation 
$\Omega^{I=0}=0$ lead to
\begin{equation}
\lambda^0_i = - {\alpha\over3\calN}\Omega_i,
\end{equation}
so that $\Omega_i=\calN^I\Omega_{Ii}$ is now essentially the $I=0$ 
component of $\lambda^I_i$. 

We have $\calA^{\bar\alpha}_i\zeta_a\equiv\zeta_i=-\Omega_i$ on shell, implying that 
the hypermultiplet fermions $\zeta_\alpha$ contain the $\Omega_i$ degree of freedom. 
To separate it out, we define new hypermultiplet fermions $\xi_\alpha$ by
\begin{equation}
\xi_\alpha\equiv\zeta_\alpha-{\calA_\alpha^i\over\calN}\Omega_i\,.
\label{eq:newhyp}
\end{equation}
Then, $\xi_\alpha$ is indeed orthogonal to $\calA^{\bar\alpha}_i$ on-shell:
\begin{equation}
\calA^{\bar \alpha}_i\xi_\alpha 
= \zeta_i-{\calA^2\over2\calN}\Omega_i
= (\zeta_i+\Omega_i)-{1\over2\calN}\Omega_i(\calA^2+2\calN) \,.
\label{eq:Axi}
\end{equation}
In the Lagrangian, the quadratic terms of the form $\bar\zeta^{\bar\alpha}\Gamma\zeta_
\alpha$ yield `cross terms' proportional to $\calA^{\bar\alpha}_i\xi_\alpha$, which
do not vanish but can be eliminated by further shifts of the multiplier 
auxiliary fields $\chi$ and $C$. Explicitly, we have
\begin{eqnarray}
\bar\zeta^{\bar\alpha}\slash{\nabla}\zeta_\alpha&=&(\bar\zeta^{\bar\alpha}\slash{\nabla}\zeta_\alpha)'
+\Bigl\{{1\over\calN}\bigl(
e^{-1}\nabla_\mu(ee^\mu_a)\,\bar\Omega^i+2{\nabla_a}\bar\Omega^i\bigr)\gamma^a(\zeta_i+\Omega_i) \nn
&&\hspace{13em}{}+{1\over2\calN^2}\bar\Omega\slash{\nabla}\Omega(\calA^2+2\calN)
\Bigr\}\,,\nn
\bar\zeta^{\bar\alpha}\Gamma\zeta_\alpha&=&(\bar\zeta^{\bar\alpha}\Gamma\zeta_\alpha)'
-\Bigl\{{2\over\calN}\bar\Omega^i\Gamma(\zeta_i+\Omega_i)
-{1\over2\calN^2}\bar\Omega\Gamma\Omega(\calA^2+2\calN)\Bigr\}\,,
\end{eqnarray}
up to a total derivative term in the action, 
where the primed terms are the `diagonal' parts:
\begin{eqnarray}
&&(\bar\zeta^{\bar \alpha}\slash{\nabla}\zeta_\alpha)'
\equiv\bar\xi^{\bar\alpha}\slash{\nabla}\xi_\alpha+{1\over\calN}\bar\Omega\slash{\nabla}\Omega 
+{1\over\calN^2}(\bar\Omega^i\gamma^a\Omega^j)\calA^{\bar\alpha}_i\nabla_a\calA_{\alpha j}
+{2\over\calN}(\bar\xi^{\bar\alpha}\gamma^a\Omega_i)\nabla_a\calA^i_\alpha\,,\nn
&&(\bar\zeta^{\bar \alpha}\gamma_{ab}\zeta_\alpha)'\equiv 
\bar\xi^{\bar \alpha}\gamma_{ab}\xi_\alpha 
+{1\over\calN}\bar\Omega\gamma_{ab}\Omega\,.
\end{eqnarray}
Collecting all the contributions from the bilinear terms in $\zeta_\alpha$, we 
find that the cross terms can be 
eliminated by replacing $C\fprm$ and $\chi\fprm$ by the shifted 
quantities $C\sprm$ and $\chi\sprm$ 
defined as
\begin{eqnarray}
\hspace*{-5mm}
C\sprm&=&C\fprm +{1\over2\calN^2}\Bigl\{-2i\bar\Omega\slash{\nabla}\Omega 
+\left(
\bar\psi_a\gamma_b\psi_c(\bar\Omega\gamma^{abc}\Omega)
-\half\bar\psi^a\gamma_{bc}\psi_a(\bar\Omega\gamma^{bc}\Omega)\right) \nn
\hspace*{-5mm}
&&\qquad \qquad \qquad {}-i\bar\Omega\gamma\dt(v-\myfrac1{2\alpha}F(A))\Omega\Bigr\}
+{i\over\calN}e^{-1}\nabla_\mu(e\bar\psi_a\gamma^\mu\gamma^a\Omega)\,, \nn
\hspace*{-5mm}
\chi\sprm_i&=&\chi\fprm_i +{1\over4\calN}\Bigl\{\left(
e^{-1}\nabla_\mu(ee^\mu_a)\,\gamma^a\Omega_i+2\slash{\nabla}\Omega_i\right) \nn
\hspace*{-5mm}
&&{}+i\left(
\gamma^{abc}\Omega_i(\bar\psi_a\gamma_b\psi_c)
-\half\gamma^{bc}\Omega_i(\bar\psi^a\gamma_{bc}\psi_a)\right) 
+\gamma\dt(v-\myfrac1{2\alpha}F(A))\Omega_i\Bigr\}\,.
\hspace{2em}
\end{eqnarray}
Here, in the last term of $C\sprm$, we have also added a contribution 
from the term 
$-4i\bar\psi^i_a\gamma^b\gamma^a\zeta_\alpha\nabla_b\calA^{\bar \alpha}_i$
in ${\cal L}'_{\rm hyper}$, which yields a term proportional to 
$\calA^2+2\calN$ after partial integration when $\zeta_\alpha$ is rewritten 
by using Eq.~(\ref{eq:newhyp}). 

We now rewrite the action (\ref{eq:actionI}) by using the 
field redefinitions (\ref{eq:newRS}), (\ref{eq:gaugino}) and 
(\ref{eq:newhyp}) everywhere. 
From this point, {\it the Rarita-Schwinger field always stands for the 
new variable $\psi_\mu^{\rm N}$, and we {\it omit} the cumbersome 
superscript} N.

Rewriting (\ref{eq:actionI}) actually involves a very tedious 
computation. Note, for instance, that the spin connection 
$\omega_\mu^{ab}|_{b_\mu=0}$ contained in the covariant derivative 
$\nabla_\mu$ and $\calR(M)$ is given in Eq.~(\ref{eq:connection}) in 
terms of the original Rarita-Schwinger field $\psi_\mu$, which should 
also be rewritten in terms of the new variable $\psi^{\rm N}_\mu$ in 
Eq.~(\ref{eq:newRS}). Surprisingly, however, all the terms containing 
$\Omega_i\equiv\calN^I\Omega_{Ii}$ completely cancel out in the action
if the auxiliary fields are eliminated by the equations of motion. 
This action, which is obtained by eliminating the auxiliary fields,
is just the action in the on-shell formulation, which we term the 
`on-shell action'. Since $\Omega_i\propto 
\lambda^{I=0}_i$, as noted above, this fact that the $\Omega_i$ 
completely disappear is the {\it fermionic counterpart} of the previously 
observed fact that the $M^{I=0}=\alpha$ and $F_{\mu\nu} 
^{I=0}(W)=F_{\mu\nu}(A)$ terms disappeared from the action. That is, 
there appear no terms that carry an explicit $I=0$ index, and the upper 
indices $I, J$, etc., are always contracted with the lower indices of 
$\calN_I, \calN_{IJ}$, etc., in the on-shell action. 

We can demonstrate this noteworthy fact as follows. 
First, we can confirm that the 
index $I$ is `conserved' in all the supersymmetry transformation laws of
the physical fields (fields other than the auxiliary fields); that is, 
the supersymmetry transformation of a physical field with the
index $I$ contains only the terms carrying the same index, and that of a
physical field without the index $I$ contains only the terms carrying no 
index. Thus the fields $\Omega_i$, $\alpha$ and $F_{\mu\nu}(A)$, carrying
the $I=0$ index explicitly, appear only in the transformation of those 
$I=0$ fields. This can be confirmed relatively easily, as we see in the 
next section. Therefore, if such terms carrying the $I=0$ index 
explicitly remain in the on-shell action, the supersymmetry invariance 
of the action implies that the parts of the action containing different 
numbers of $I=0$ fields are separately supersymmetry invariant. But we 
know already that the bosonic $I=0$ fields $\alpha$ and $F_{\mu\nu}(A)$ 
do not appear. Clearly, no such invariant term can be made from the $\Omega_i$ 
without using their superpartners $\alpha$ and $F_{\mu\nu}(A)$. This 
proves the total cancellation of the $\Omega_i$ terms in the on-shell 
action. (We have also confirmed this cancellation explicitly by direct 
rewriting of the action, except for some four-fermion term parts.) 

Completing the square of the auxiliary field terms in the action 
(\ref{eq:actionI}), we can rewrite the action in a sum of the 
on-shell action and the perfect square terms of the auxiliary fields.
The auxiliary fields implicitly contain $\Omega_i$-dependent terms in them. 
This can be seen by substituting the field redefinitions (\ref{eq:newRS}), 
(\ref{eq:gaugino}) and (\ref{eq:newhyp}) into their solutions of the 
equations of motion. 
If we redefine the auxiliary fields as follows by 
subtracting these implicitly $\Omega_i$-dependent terms,
then the $\Omega_i$-dependent terms completely 
disappear also from the perfect square terms of the auxiliary fields, 
and we have
\begin{eqnarray}
\tV^{ij}_a&=& V^{ij}_a +
{1\over2\calN}\bigl(4i\bar\Omega^{(i}\psi_a^{j)}
+{2i\over3\calN}\bar\Omega^i\gamma_a\Omega^j\bigr), \nn
\tv_{ab} &=& v_{ab}-{1\over2\alpha}F_{ab}(A) +i\bar\psi_a\psi_b 
+ i{2\over3\calN}\bar\psi_{[a}\gamma_{b]}\Omega+{i\over9\calN^2}\bar
\Omega\gamma_{ab}\Omega,\nn
\ty^{Iij} &=&\calP^I_JY^{Jij}-{2i\over3\calN}\bar\lambda^{I(i}\Omega^{j)}\,, 
\nn
\ttp^{ij} &=& t^{ij} -{\calN_IY^{Iij}\over3\calN}
        +{i\over9\calN^2}\bar\Omega^{(i}\Omega^{j)}, 
\end{eqnarray}
where 
$\calP^I_J$ is the projection operator introduced in 
Eq.~(\ref{eq:Proj}), and we have taken into account the fact that 
$Y^I-M^It=\calP^I_JY^J-M^I(t-\calN_JY^J/3\calN)$. Note that the vector 
multiplet auxiliary fields $\ty^I$ as well as $\calP^I_JY^J$ are 
orthogonal to $\calN_I$, as are the fermionic partners $\lambda^I$. The 
solutions of the equations of motion for these auxiliary fields are now 
free from the $\Omega_i$ and given by
\begin{eqnarray}
\tV_{{\rm sol}\,a}^{ij}&=& 
-{1\over2\calN}\bigl(2\calA^{\bar\alpha(i}\nabla_a\calA_\alpha^{j)}
-i\calN_{IJ}\bar\lambda^{Ii}\gamma_a\lambda^{Jj}\bigr), \nn
\tv_{{\rm sol}\,ab}&=& 
-{1\over4\calN}\Bigl\{
\calN_I F_{ab}(W)^I 
-i\left(6\calN\bar\psi_a\psi_b+\bar\xi^{\bar \alpha}\gamma_{ab}\xi_\alpha 
-\half\calN_{IJ}\bar\lambda^I\gamma_{ab}\lambda^J\right)\Bigr\}, \nn
\ty^{Iij}_{\rm sol}&=&-\half a^{IJ}\calP^K_J\calY_K^{ij}
=-\half\calP^I_Ja^{JK}\calY_K^{ij}
=-\left(\half a^{IJ}-\myfrac13M^IM^J\right)\calY_J^{ij} \nn
&&\hbox{with}\quad  \calY_I^{ij}\equiv 
2\calA^{(i}_{\alpha}(gt_I)^{\bar\alpha\beta}\calA_\beta^{j)}
+i\calN_{IJK}\bar\lambda^{Ji} \lambda^{Kj}, \nn
\ttp_{\rm sol}^{ij}&=& -{1\over6\calN}M^I\calY_I^{ij}
=-{1\over6\calN}
\bigl(2\calA^{(i}_{\alpha}(gM)^{\bar\alpha\beta}\calA_\beta^{j)}
+i\calN_{IJ}\bar\lambda^{Ii} \lambda^{Jj}\bigr), 
\label{eq:sol}
\end{eqnarray}
where $a^{IJ}$ is the inverse of the metric $a_{IJ}$ of the 
vector multiplet kinetic terms:
\begin{equation}
a_{IJ}\equiv-\half{\partial^2\over\partial M^I\partial M^J}\ln\calN=-{1\over2\calN}
\bigl(\calN_{IJ}-{\calN_I\calN_J\over\calN}\bigr), \quad 
a^{IJ}\equiv(a^{-1})^{IJ}.
\end{equation}
It possesses the properties
\begin{eqnarray}
a_{IJ}M^J=\calN_I/2\calN \ \rightarrow\ a^{IJ}\calN_J/2\calN=M^I,\qquad 
a^{IJ}\calP^K_J=\calP^I_Ja^{JK}.
\end{eqnarray}
We here have assumed that $a_{IJ}$ is invertible. However, 
there are some interesting cases in which $\det(a_{IJ})=0$. 
Such a situation implies that some vector multiplets have no kinetic terms, 
since $a_{IJ}$ gives the metric of the vector multiplets. We 
comment on such a possibility below.

After all of the above calculations,
 the action is finally found to take the form
\begin{eqnarray}
&&\hspace{-5.5em}{\cal L}
={\cal L}_{\rm hyper}+{\cal L}_{\rm vector}+{\cal L}_{\rm C\hbox{-}S}+{\cal L}_{\rm aux}\ , \nn
e^{-1}{\cal L}_{\rm hyper}&=&\nabla^a\calA_i^{\bar \alpha}\nabla_a\calA^i_\alpha 
-2i\bar\xi^{\bar \alpha}(\slash{\nabla}+gM)\xi_\alpha\nn
&&{}
+\calA^{\bar\alpha}_i(gM)^2{}_\alpha{}^\beta\calA_\beta^i
-4i\bar\psi^i_a\gamma^b\gamma^a\xi_\alpha\nabla_b\calA^{\bar \alpha}_i 
-2i\bar\psi^{(i}_a\gamma^{abc}\psi_c^{j)}\calA^{\bar\alpha}_j\nabla_b\calA_{\alpha i}\nn
&&{}+\calA^{\bar\alpha}_i
\bigl(8ig\bar\lambda^i_{\alpha\beta}\xi^\beta-4ig\bar\psi_a^i\gamma^aM_{\alpha\beta}\xi^\beta 
\nn &&{}
\qquad \qquad +4ig\bar\psi_a^{(i}\gamma^a\lambda^{j)}_{\alpha\beta}\calA^\beta_j 
-2ig\bar\psi^{(i}_a\gamma^{ab}\psi^{j)}_bM_{\alpha\beta}\calA^\beta_j\bigr)
\nn
&&{}+\bar\psi_a\gamma_b\psi_c\bar\xi^{\bar\alpha}\gamma^{abc}\xi_\alpha 
-\myfrac12\bar\psi^a\gamma^{bc}\psi_a\bar\xi^{\bar\alpha}\gamma_{bc}\xi_\alpha\ ,\nn
e^{-1}{\cal L}_{\rm vector}&=&
  -\half R(\omega)-2i\bar\psi_\mu\gamma^{\mu\nu\rho}\nabla_\nu\psi_\rho 
  +(\bar\psi_a\psi_b)(\bar\psi_c\gamma^{abcd}\psi_d+\bar\psi^a\psi^b) \nn
&&{}-\calN_I \left(
   ig[\bar\lambda,\,\lambda]^I - \myfrac{i}4\bar\psi_c\gamma^{abcd}\psi_dF_{ab}(W)^I
\right)\nn
&&{}+ a_{IJ}
\left(\begin{array}{c}
-\myfrac14F(W)^I\dt F(W)^J+\myfrac12\nabla_aM^I\nabla^aM^J \\[.6ex]
+2i\bar\lambda^I\slash{\nabla}\lambda^J 
+i\bar\psi_a(\gamma\dt F(W)-2\slash{\nabla}M)^I\gamma^a\lambda^J \\[.5ex]
-2(\bar\lambda^{I}\gamma^a\gamma^{bc}\psi_a)(\bar\psi_b\gamma_c\lambda^{J})
+2(\bar\lambda^{I}\gamma^a\gamma^{b}\psi_a)(\bar\psi_b\lambda^{J})
\end{array}\right) \nn
&&{}-\calN_{IJK}\left(\begin{array}{c}
-i\bar\lambda^I\myfrac14\gamma\dt F(W)^J\lambda^K \\[.4ex]
{}+\myfrac23(\bar\lambda^I\gamma^{ab}\lambda^{J})(\bar\psi_a\gamma_b\lambda^{K}) 
+\myfrac23(\bar\psi^i\dt\gamma\lambda^{Ij})(\bar\lambda_{(i}^J\lambda_{j)}^K) 
\end{array}\right)\nn
&&
+\myfrac18\left(2\bar\psi_a\psi_b
+\bar\xi^{\bar \alpha}\gamma_{ab}\xi_\alpha 
+a_{IJ}\bar\lambda^I\gamma_{ab}\lambda^J\right)^2 \nn
&&{}+i\myfrac14\calN_IF(W)^I\left(2\bar\psi_a\psi_b
+\bar\xi^{\bar \alpha}\gamma_{ab}\xi_\alpha 
+a_{IJ}\bar\lambda^I\gamma_{ab}\lambda^J\right) \nn
&&{}+\bigl(\calA^{\bar\alpha i}\nabla_a\calA_\alpha^j
+ia_{IJ}\bar\lambda^{Ii}\gamma_a\lambda^{Jj}\bigr)^2 \nn
&&{}-\myfrac14(a^{IJ}-M^IM^J)
\calY_I^{ij}\calY_{J\,ij}\ .
\label{eq:finalAction}
\end{eqnarray}
Here ${\cal L}_{\rm aux}$ represents the perfect square terms of 
the auxiliary fields, which vanish on shell:
\begin{eqnarray}
e^{-1}{\cal L}_{\rm aux}&=&
C\tprm(\calA^2+2) -8i\bar\chi\sprm{}^i\calA^{\bar\alpha}_i\xi_\alpha\nn
&&{}+2(\tv-\tv_{\rm sol})^2 
-(\tV-\tV_{\rm sol})^{ij}(\tV-\tV_{\rm sol})_{ij} \nn
&&{}-3(\ttp-\ttp_{\rm sol})^{ij}(\ttp-\ttp_{\rm sol})_{ij}
+a_{IJ}(\ty^I-\ty^I_{\rm sol})^{ij}(\ty^J-\ty^J_{\rm sol})_{ij} \nn
&&{}+\left(1-{A^2/\alpha^2}\right)
(\calF^{\bar\alpha}_i-\calF^{\bar\alpha}_{{\rm sol}\,i})
(\calF^i_\alpha-\calF^{\,i}_{{\rm sol}\,\alpha})\ .
\label{eq:Laux}
\end{eqnarray}
Here the multiplier term $C\sprm(\calA^2+2\calN) -8i\bar\chi\sprm(\zeta+\Omega)$ has 
been rewritten into the form of the first line by using Eq.~(\ref{eq:Axi})
and defining $C\tprm$ in terms of the $C\sprm$ field as
\begin{equation}
C\tprm=C\sprm-i{4\over\calN}\bar\chi\sprm\Omega\,.
\end{equation}
Expressed in this way, the explicit $\Omega_i$ have been completely 
removed from
the action. Note that the final action (\ref{eq:finalAction}) with 
(\ref{eq:Laux}) is everywhere written in terms of the new variables, 
although the superscript N has been omitted. In particular, the spin 
connection $\omega_\mu^{ab}$ in the covariant derivative $\nabla_\mu$ and $R(\omega)$ 
is the new one given by Eq.~(\ref{eq:connection}) with the new $\psi_\mu$ 
used and $b_\mu$ set equal to 0. By using this $\omega_\mu^{ab}$, \ $R(\omega)$ is 
given as usual: 
\begin{equation}
R_{\mu\nu}{}^{ab}(\omega)=
2\partial_{[\mu}\omega_{\nu]}{}^{ab}-2\omega_{[\mu}{}^{[ac}\omega_{\nu]c}{}^{b]}, \quad 
R_{ab}(\omega)\equiv R_{ac}{}^c{}_b(\omega)\,,
\quad R(\omega)\equiv R_a{}^a(\omega)\,. 
\end{equation}

\section{Supersymmetry transformation}

Now we should modify the supersymmetry ($\XQ$) transformation 
$\delta_Q(\varepsilon)$, since 
we have fixed the $\XD$ gauge by (\ref{eq:Dgauge}) and made various 
field redefinitions, (\ref{eq:newRS}), 
(\ref{eq:gaugino}) and (\ref{eq:newhyp}).
The proper $\XQ$ transformation is found to be given by the following 
linear combination of the original transformations of $\XQ$, 
dilatation $\XD$, local-Lorentz $\XM$ and $SU(2)$ $\XU$:
\begin{eqnarray}
\delta^{\rm N}_Q(\varepsilon)&=&\delta_Q(\varepsilon)+\delta_D(\rho(\varepsilon))
+\delta_M(\lambda^{ab}(\varepsilon))+\delta_U(\theta^{ij}(\varepsilon)),\nn
   \rho(\varepsilon)&\equiv&-\myfrac{2i}{3\calN}\bar\varepsilon\Omega,\quad 
  \lambda^{ab}(\varepsilon)\equiv\myfrac{2i}{3\calN}\bar\varepsilon\gamma^{ab}\Omega,\quad 
\theta^{ij}(\varepsilon)\equiv-\myfrac{2i}{\calN}\bar\varepsilon^{(i}\Omega^{j)} .
\end{eqnarray}
The dilatation part $\delta_D(\rho(\varepsilon))$ is determined so as to
maintain the $\XD$ gauge fixing condition (\ref{eq:Dgauge}): 
$\left(\delta_Q(\varepsilon)+\delta_D(\rho(\varepsilon))\right)\calN=0$.
The local-Lorentz part $\delta_M(\lambda^{ab}(\varepsilon))$ is fixed by
requiring that the transformation of the f\"unfbein take the canonical 
form $\delta^{\rm N}(\varepsilon)e_\mu{}^a 
=-2i\bar\varepsilon\gamma^a\psi^{\rm N}_\mu$ in terms of the {\it new} 
Rarita-Schwinger field $\psi^{\rm N}_\mu$. In the first part of this 
section, we revive the superscript N to distinguish the new variables 
from the original ones. Finally, the $SU(2)$ part $\delta_U(\theta^{ij})$ 
is added so that the hypermultiplet scalar field $\calA^i_\alpha$ is 
transformed in the new fermion component $\xi_\alpha$ to yield the form 
$\delta^{\rm N}(\varepsilon)\calA^i_\alpha 
=2i\bar\varepsilon^i\xi_\alpha$.

To write the supersymmetry transformation rules concisely and covariantly, 
we should use the supercovariant derivative $\hatD_\mu$
and the supercovariantized curvatures $\hatR_{\mu\nu}$. 
But these supercovariant quantities are also modified by the $\XD$ gauge
fixing and field redefinitions. 
We define a new supercovariant derivative $\hatD^{\rm N}_\mu$ in the 
usual form, but by using the new gauge fields and the new 
supersymmetry transformation:
\begin{equation}
\hatD^{\rm N}_\mu=\partial_\mu-\delta_M(\omega^{{\rm N}\,ab}_\mu)-\delta_U(\tV_\mu^{ij})
-\delta_G(W_\mu)-\delta_Q^{\rm N}(\psi^{\rm N}_\mu).
\end{equation}
The relation with the original supercovariant derivative $\hatD_\mu$, 
which also contains the $\XD$ covariantization, is found to be given by
\begin{equation}
\hatD_\mu=\hatD_\mu^{\rm N}
-\delta_D(b^{\rm N}_\mu)+\delta_M\left(2e_\mu{}^{[a}\N b^{b]}
+\myfrac{i}{9\calN^2}\bar\Omega\gamma_\mu{}^{ab}\Omega\right)
-\delta_U(\myfrac{i}{3\calN^2}\bar\Omega^{(i}\gamma_\mu\Omega^{j)})
-\delta^{\rm N}_Q(\myfrac1{3\calN}\gamma_\mu\Omega) ,
\label{eq:Drel} 
\end{equation}

\vspace*{3mm}
\noindent
where $b^{\rm N}_\mu$ is the supercovariantized $b_\mu$ 
$(=\alpha^{-1}\partial_\mu\alpha)$ defined as
$b^{\rm N}_\mu\equiv\alpha^{-1}\hatD^{\rm N}_\mu\alpha 
=b_\mu+\myfrac{2i}{3\alpha\calN}\bar\psi^{\rm N}_\mu\Omega.$ 
The new curvatures $\hatR^{\rm N}_{ab}{}^{\bar A}$
are defined as usual by Eq.~(I\,2$\cdot$28) by using the new covariant 
derivative 
$\hatD^{\rm N}_a$ with flat index $a$: 
$[\hatD^{\rm N}_a,\,\hatD^{\rm N}_b] 
= -\hatR^{\rm N}_{ab}{}^{\bar A}\XX_{\bar A}$.
Hence, using the relation (\ref{eq:Drel}) between 
$\hatD_\mu$ and $\hatD^{\rm N}_\mu$, we can find the relations between 
the new curvatures and the original curvatures $\hatR_{ab}{}^{\bar A}$. 
The Yang-Mills group $G$ is also regarded as a subgroup of our supergroup,
and so, for example, in the case $\bar A=I$ of $G$, we find
\begin{equation}
\hat F^I_{ab}(W)={\hat F}^{{\rm N}I}_{ab}(W)
+\myfrac{4i}{3\calN}\bar\Omega\gamma_{ab}\lambda^I
+\myfrac{2i}{9\calN^2}M^I\bar\Omega\gamma_{ab}\Omega\,.
\end{equation}

From this point, we again suppress the cumbersome superscript N of 
$\psi_\mu ^{\rm N}$, $\omega_\mu^{{\rm N}\,ab}$, $\hatD^{\rm N}_\mu$, 
$\hatR^{\rm N}_{\mu\nu}{}^A$ ($\hat F^{{\rm N}I}_{\mu\nu}$) and 
$\delta^{\rm N}_Q(\varepsilon)$, since {\it every quantity that appears
in the following is always one of these new ones.}

As mentioned in the preceding section, we find that the (new) 
supersymmetry transformation `conserves' the index $I$, and thus the 
$\Omega_i\propto\lambda_i^{I=0}$, as well as $F_{ab}(A)=F_{ab}^{I=0}(W)$ 
and $\alpha=M^{I=0}$ (or $b_\mu=\alpha^{-1}\partial_\mu\alpha$), 
carrying an $I=0$ index, do not explicitly appear in the transformation 
laws, unless the transformed field itself carries $I=0$. (The only 
exception is the transformation $\delta\calF_\alpha^i$, which contains 
$A_\mu=W_\mu^{I=0}$ and $\alpha=M^{I=0}$ explicitly. However, 
$\calF^i_\alpha$ is defined to be $\delta_Z(\alpha)\calA^i_\alpha$ with 
$\alpha=M^{I=0}$, and so it may be regarded as carrying the index $I=0$ 
implicitly.) \ It is quite easy to demonstrate the disappearance of 
$F_{ab}(A)=F_{ab}^{I=0}(W)$ and $\alpha=M^{I=0}$ by direct computation. 

To see the disappearance of explicit $\Omega_i$ factors, however, we 
have proceeded in the following way. For the physical fields, 
$e_\mu{}^a, \psi_\mu^i, W_\mu^I, M^I, \lambda^I, \calA^i_\alpha, \xi_
\alpha$, we have explicitly computed their supersymmetry transformation laws
and directly checked that the explicit $\Omega_i$ cancel out completely. For
the auxiliary fields $\phi=\tV_\mu^{ij}, \ttp^{ij}, \tv_{ab}, \ty^{ij}, 
\calF^i_\alpha$, other than $\chi\sprm$ and $C\tprm$, such rigorous 
computations become quite tedious, and so we checked the cancellation 
indirectly: For 
such auxiliary fields $\phi$, the supersymmetry transformation of 
$\phi-\phi_{\rm sol}$, $\delta(\phi-\phi_{\rm sol})$, should vanish 
on-shell, that is, when the equations of motion for auxiliary fields are
used. (But note that the equations of motion for the physical fields 
need not be used.) Therefore, if an $\Omega_i$ appears explicitly in 
$\delta(\phi-\phi_{\rm sol})$, it must be multiplied by the factors 
$(\phi-\phi_{\rm sol})$ which vanish on-shell, or it must appear in
the form $\Omega_i+\zeta_i$. For the former possibility, we can easily 
see whether such terms appear or not, by keeping track of auxiliary 
fields explicitly. It is seen that the
 latter possibility does not  occur by 
confirming that $\zeta_i=\calA^{\bar\alpha}_i\zeta_\alpha$ never appears
in $\delta(\phi-\phi_{\rm sol})$. Once $\Omega_i$ is seen to be absent 
in $\delta (\phi-\phi_{\rm sol})$, it is seen that it does not appear in
$\delta\phi$ either, since $\phi_{\rm sol}$ consists of physical fields 
alone, and hence $\delta\phi_{\rm sol}$ does not contain any $\Omega_i$
explicitly. 

Computations to derive transformation laws of the new auxiliary fields 
$\chi\sprm$ and $C\tprm$ directly from those of the original fields 
$\chi$ and $C$ become terribly tedious, because the relations between 
these new and original fields are very complicated. Instead of doing 
this, we can use the invariance of the action to find 
$\delta\chi\sprm$ and $\delta C\tprm$. Then, since they appear in the 
form $\delta C\tprm\,(\calA^2+2\calN) 
-8i\delta\bar\chi\sprm^i\,\calA^{\bar\alpha}_i
\xi_\alpha$ in $\delta{\cal L}$, we have only to compute the terms whose
supersymmetry transformations can yield the factor $\calA^2$ or 
$\calA^{\bar\alpha}_i\xi_\alpha $. There are not a great number of such 
terms in the action. The cancellation condition for the terms 
proportional to $(\calA^2+2\calN)$ and $\calA^{\bar\alpha}_i\xi_\alpha$ 
determines the supersymmetry transformation laws $\delta C\tprm$ and 
$\delta\chi\sprm$ as follows:
\begin{eqnarray}
\delta\pr \chi^i&=&\half\varepsilon^iC\tprm
+\pr \chi^i(2i\bar\varepsilon\gamma\dt \psi)-\psi_a^j(2i\bar\varepsilon_j\gamma^a\pr \chi^i)
-\half\slash\nabla({\mbf\Gamma}\varepsilon^i)\nn
&&+\myfrac{i}2\gamma^a{\mbf\Gamma}\varepsilon^i(\bar\psi_a\gamma\dt \psi)
-\myfrac{i}4\gamma^{abc}{\mbf\Gamma}\varepsilon^i(\bar\psi_a\gamma_b\psi_c)
+\myfrac{i}8\gamma^{ab}{\mbf\Gamma}\varepsilon^i(\bar\psi^c\gamma_{ab}\psi_c+2\bar\psi_a\psi_b)\nn
&&-\myfrac14\gamma\dt \tilde v\,{\mbf\Gamma}\varepsilon^i
-\half e^{-1}\nabla_\lambda(e\tilde V^\lambda)\varepsilon^i
+\myfrac{i}2e^{-1}\nabla_\lambda(e\bar\psi_\mu^i\gamma^{\mu\lambda\nu}\psi_\nu^j)\varepsilon_j\,,\nn
\delta C\tprm&=&-2ie^{-1}\nabla_\mu(e\bar\varepsilon\gamma^\mu\pr\chi)
-ie^{-1}\nabla_\mu(e\bar\psi_a\gamma^\mu\gamma^a{\mbf\Gamma}\varepsilon) \nn
&&{}+C\tprm2i\bar\varepsilon\gamma\dt \psi+4i\bar{\pr\chi}{\mbf\Gamma}\varepsilon\,.
\label{eq:TrfChiC}
\end{eqnarray}
Here ${\mbf\Gamma}$ is a field-dependent matrix acting on a spinor with an 
$SU(2)$ index which is defined by
\begin{equation}
{\mbf\Gamma}\varepsilon^i\equiv(-\gamma\dt{\tilde V}+3\tilde t)^i{}_j\varepsilon^j
+\gamma\dt \tilde v\varepsilon^i+\myfrac{\calN_I}{4\calN}\gamma\dt{\hat F}^I(W)\varepsilon^i
+\myfrac{\calN_{IJ}}{\calN}\lambda^{iI}(2i\bar\lambda^J\varepsilon)\,.\nn
\label{eq:Psi}
\end{equation}
Note that there appear derivative terms of the transformation
parameter, $\partial_\mu\varepsilon^i$, in these, implying 
that $\chi\sprm$ and $C\tprm$ are not covariant quantities.
For this reason we redefine these fields once again as follows by adding
proper supercovariantization terms:
\begin{eqnarray}
\tilde\chi^i\equiv\pr \chi^i+\half\gamma^a{\mbf\Gamma}\psi^i_a\,,\qquad 
\tilde C\equiv C\tprm+2i {\bar\psi}\dt\gamma\tilde\chi 
-i {\bar\psi}_a\gamma^{ab}{\mbf\Gamma}\psi_b\,.
\end{eqnarray}
Here we have used 
the identity $(\nabla_\mu\bar\varepsilon){\mbf\Gamma}\psi_a=\bar\psi_a{\mbf\Gamma}\nabla_\mu\varepsilon$
in deriving the covariantization terms for $C\tprm$. 

We must next derive the supersymmetry transformation law for these covariant 
variables $\tilde\chi$ and $\tilde C$ from Eq.~(\ref{eq:TrfChiC}). 
Note here the simple fact that the transformation of any covariant 
quantity gives a covariant quantity and hence cannot contain gauge 
fields explicitly; that is, gauge fields can appear only implicitly in 
the covariant derivatives or in the form of supercovariant curvatures 
(field strengths). Otherwise, the two sides of the commutation relation of 
the transformations would lead to a contradiction. This observation 
greatly simplifies the computations of $\delta\tilde\chi$ and 
$\delta\tilde C$, in which we can discard such explicit gauge field 
terms, since they are guaranteed to cancel out anyway.

We now write the final supersymmetry transformation laws derived this way. 
The $\XQ$ transformation laws  of the Weyl multiplet are
\begin{eqnarray}
\delta e_\mu{}^a&=&-2i\bar\varepsilon\gamma^a\psi_\mu\,,\nn
\delta\psi_\mu^i&=& {\calD}_\mu\varepsilon^i+\gamma_\mu\tilde t^i{}_j\varepsilon^j
+\half\gamma_{\mu ab}\varepsilon^i\tilde v^{ab}
+\myfrac{\calN_I}{12\calN}\gamma_\mu\gamma\dt {\hat F}^I(W)\varepsilon^i
+\myfrac{\calN_{IJ}}{3\calN}\gamma_\mu\lambda^{Ii}(2i\bar\lambda^J\varepsilon)\,,\nn
\delta\tilde V_\mu^{ij}&=&-4i\bar\varepsilon^{(i}\gamma_\mu\tilde\chi^{j)}-i\bar\varepsilon^{(i}\gamma_{\mu 
ab} \hatR{}^{abj)}(Q)
+4i\bar\varepsilon^{(i}\gamma\dt \left(\tilde v
 +\myfrac{\calN_I}{4\calN} {\hat F}^I(W)\right) \psi{}_\mu^{j)} \nn
&&{}-6i(\bar\varepsilon{ \psi}_\mu)\ttp^{ij}+\myfrac{4\calN_{IJ}}{\calN}
\left( ({\bar\psi}_\mu\lambda^I)\bar\varepsilon^{(i}\lambda^{j)J}
-(\bar\varepsilon\lambda^I) {\bar\psi}{}_\mu^{(i}\lambda^{j)J}\right),\nn
\delta\tilde t^{ij}
&=&4i\bar\varepsilon^{(i}\tilde\chi^{j)}+i\bar\varepsilon^{(i}\gamma\dt  \hatR{}^{j)}(Q)
+\myfrac{2i\calN_{IJ}}{3\calN}\left(\bar\varepsilon^{(i} \slashD M^I\lambda^{j)J}
-(\bar\varepsilon\lambda^I)\tilde Y^{Jij}\right),\nn
\delta\tilde v_{ab}&=&-2i\bar\varepsilon\gamma_{ab}\tilde\chi 
-\myfrac{i}2\bar\varepsilon\gamma_{ab}\gamma\dt  \hatR(Q)
+\myfrac{i}4\bar\varepsilon\gamma_{abcd} \hatR{}^{cd}(Q)-i\bar\varepsilon\hatR_{ab}(Q)\,,\nn
\delta\tilde\chi^i&=&\half\varepsilon^i\tilde C-\half( {\slashD}{\mbf\Gamma}')\varepsilon^i
-\myfrac14\gamma\dt\tilde v\,{\mbf\Gamma}'\varepsilon^i+\myfrac14\gamma\dt {\hatR}(U)\varepsilon^i \nn
&&+\half\gamma^a{\mbf\Gamma}'\left(\gamma_a\tilde t\varepsilon^i+\half\gamma_{abc}\varepsilon^i\tilde v^{bc}
+\myfrac{\calN_I}{12\calN}\gamma_a\gamma\dt {\hat F}^I(W)\varepsilon^i
-\myfrac{\calN_{IJ}}{3\calN}\gamma_a\lambda^{Ii}(2i\bar\varepsilon\lambda^J)\right),\nn
\delta\tilde C&=&-2i\bar\varepsilon{\slashD}\tilde\chi 
+\half i\bar\varepsilon\{ \gamma^{ab},\,{\mbf\Gamma}'\}{\hatR}_{ab}(Q)
+\myfrac{2i}3\bar\varepsilon{\mbf\Gamma}'\tilde\chi+\myfrac{i}{3}\bar\varepsilon\gamma 
\dt\tilde v\tilde\chi\,,\nn
\delta\omega_\mu{}^{ab}&=&-2i\bar\varepsilon\gamma^{[a} {\hatR}_\mu{}^{b]}(Q)
-i\bar\varepsilon\gamma_\mu\hatR{}^{ab}(Q)\nn
&&{}-2i\bar\varepsilon\gamma^{abcd}\psi_\mu 
\left(\tilde v_{cd}+\myfrac{\calN_I}{6\calN}{\hat F}^I_{ab}(W)\right)
+2i\bar\varepsilon\psi_\mu\myfrac{\calN_I}{3\calN}{\hat F}^I_{ab}(W)\nn
&&{}-4i\bar\varepsilon^i\gamma^{ab}\psi_\mu^j\tilde t_{ij}
+\myfrac{4\calN_{IJ}}{3\calN}\left((\bar\varepsilon\lambda^I){\bar\psi}_\mu\gamma^{ab}\lambda^J
-(\bar\varepsilon\gamma^{ab}\lambda^I){\bar\psi}_\mu\lambda^J\right), 
\label{eq:gaugeTrf}
\end{eqnarray}
where $\calD_\mu$ is the covariant derivative that is covariant 
only with respect to homogeneous transformations $\XM_{ab},\XU^{ij}$ and
$\XG$, and the prime on ${\mbf\Gamma}$ implies that $\XU$-gauge field in 
${\mbf\Gamma}$ is removed: 
${\mbf\Gamma}\varepsilon^i={\mbf\Gamma}'\varepsilon^i-\gamma\dt{\tilde V}^i_{\,j}\varepsilon^j$. 
Here we have also written 
the transformation law of the spin connection for convenience, 
although it is a dependent field. 

The supersymmetry transformation laws of the vector multiplet are 
\begin{eqnarray}
\delta W^I_\mu&=&-2i\bar\varepsilon\gamma_\mu\lambda^I+2i\bar\varepsilon\psi_\mu M^I,\nn
\delta M^I&=&2i\bar\varepsilon\lambda^I,\nn
\delta\lambda^I_i&=&\calP^I{}_J(-\myfrac14\gamma\dt {\hat F}^J(W)\varepsilon_i
-\myfrac12\slashD M^J\varepsilon_i+\tilde Y^J_{ij}\varepsilon^j)
-\myfrac{M^I\calN_{JK}}{3\calN}2i\bar\varepsilon\lambda^J\lambda^K_i,\nn
\delta\tilde Y^{Iij}&=&2i\bar\varepsilon^{(i}\calP^I{}_J\slashD\lambda^{Jj)}
-i\bar\varepsilon^{(i}\gamma\dt \tv\lambda^{Ij)}
-i\myfrac{\calN_J}{6\calN}\bar\varepsilon^{(i}\gamma\dt {\hat F}^J(W)\lambda^{Ij)}\nn
&&{}+2i\bar\varepsilon^{(i}\tilde t^{j)}{}_k\lambda^{Ik}+4i\bar\varepsilon\lambda^I\tilde t^{ij}
-2ig\bar\varepsilon^{(i}[M,\,\lambda]^{Ij)} \nn
&&{}+\myfrac{4\calN_{JK}}{3\calN}\bar\varepsilon\lambda^J\bar\lambda^{K(i}\lambda^{Ij)}
-\myfrac{M^I\calN_{JK}}{3\calN}2i\bar\varepsilon\lambda^J\tilde Y^{Kij}.
\label{eq:vecSuper}
\end{eqnarray}

Finally, the hypermultiplet transformation laws are given by
\begin{eqnarray}
\delta\calA_\alpha^i&=&2i\bar\varepsilon^i\xi_\alpha\,,\nn
\delta\xi_\alpha 
&=&-{\slashD}\calA_\alpha^i\varepsilon_i+\varepsilon_igM_{\alpha\beta}\calA^{i\beta}+{\mbf\Gamma}'\varepsilon_i\calA^i_\alpha 
+\bigl(1+\myfrac{\slash A}\alpha\bigr)\varepsilon_i\tilde\calF^i_\alpha\,,\nn
\delta\tilde\calF_\alpha^i
&=&-2i\bar\varepsilon^i\bigl(1-\myfrac{\slash A}{\alpha}\bigr)^{-1}
\bigl({\slashD}\xi_\alpha+2\tilde\chi_j\calA^j_\alpha+gM_{\alpha\beta}\xi^\beta 
+2g\lambda_{j\alpha\beta}\calA^{j\beta} \nn 
&& \hspace{9em}{}
+\myfrac12\gamma\dt\tv\,\xi_\alpha+\myfrac2\alpha\lambda^0_j\tilde\calF^j_\alpha 
\bigr) 
+\myfrac{2i}{\alpha}\bar\varepsilon\lambda^0\tilde\calF^i_\alpha\,.
\label{eq:hypTrf}
\end{eqnarray}
Here $gM=M^Igt_I$ and $g\lambda_i=\lambda_i^Igt_I$ include the $I=0$ part with 
$gt_{I=0}$ as defined in Eq.~(\ref{eq:undstd0}), and the $G$ 
covariantization for $I=0$ in $\hatD_\mu$ is understood to be 
$-A_\mu(gt_0)$, instead of the original central charge transformation 
$-\delta_Z(A_\mu)$.
It is, however, interesting that the supersymmetry transformation rules 
for the latter two fields can be rewritten in slightly simpler forms if 
we refer to the original central charge transformation:
\begin{eqnarray}
\delta\xi_\alpha 
&=&-{\slashD}_*\calA_\alpha^i\varepsilon_i+\varepsilon_igM_{*}\calA^i_\alpha 
+{\mbf\Gamma}'\varepsilon_i\calA^i_\alpha\,,\nn
\delta\calF_\alpha^i&=&-2i\bar\varepsilon^i\left({\slashD}_*\xi_\alpha 
+2\tilde\chi_j\calA^j_\alpha+gM'_{\alpha\beta}\xi^\beta+2g\lambda'_{j\alpha\beta}\calA^{j\beta}
+\myfrac12\tilde v\xi_\alpha\right)
+\myfrac{4i}\alpha\bar\varepsilon^{(i}\lambda^{0j)}\calF_{\alpha j}\,.
\nn[-.8ex]
\end{eqnarray}
Here $\hatD_*$ and $M_*$ denote that 
the group action for the $I=0$ part is the original central charge 
transformation $\XZ$; that is,
\begin{equation}
\hatD_{*\mu} = \hatD_\mu' - \delta_Z(A_\mu), \qquad 
gM_*\phi^\alpha= gM'^\alpha_{\beta}\phi^\beta+ \delta_Z(\alpha)\phi^\alpha,
\end{equation}
and the primes on $\hatD'$, $gM'$ and $g\lambda_i'$ 
denote that the $I=0$ parts are omitted.\footnote{%
It may be worth mentioning that the transformation rules in
Eq.~(\ref{eq:hypTrf}) can also be rewritten equivalently by making the 
replacements $\tilde\calF^i_\alpha,\ \hatD,\  gM,\ g\lambda_i$ $\rightarrow$
$\calF^i_\alpha,\ \hatD',\  gM',\ g\lambda_i'$. }
 The central charge transformation 
given in Eq.~(I\,4$\cdot$5) can be rewritten in terms of our new variables, 
and reads, explicitly for $\calA^i_\alpha$ and $\xi_\alpha$, as
$\delta_Z(\alpha)\calA^i_\alpha=\calF^i_\alpha$ and
\begin{equation}
\delta_Z(\alpha)\xi_\alpha=
-\bigl({\slashD}_*\xi_\alpha 
+2\tilde\chi_j\calA^j_\alpha+gM'_{\alpha\beta}\xi^\beta+2g\lambda'_{j\alpha\beta}\calA^{j\beta}
+\myfrac12\tilde v\xi_\alpha\bigr)
-\myfrac{2}\alpha\lambda^{0}_j\calF_{\alpha}^j\,.
\end{equation}
The last equation is equivalent to the central charge property of the $\XZ$ 
transformation on $\calA^i_\alpha$, 
$0=\alpha\,[\delta_Z,\,\delta_Q(\varepsilon)]\calA^i_\alpha= 2i\bar\varepsilon^i\delta_Z(\alpha)\xi_\alpha 
-\alpha\delta_Q(\varepsilon)(\calF^i_\alpha/\alpha)$, which can also be rewritten in the 
following form, with $g\lambda_j{}_*\equiv g\lambda'_j+(\lambda^0_j/\alpha)\delta_Z(\alpha)$:
\begin{equation}
{\slashD}_*\xi_\alpha 
+2\tilde\chi_j\calA^j_\alpha+gM_*\xi_\alpha 
+2g\lambda_j{}_*\calA^j_\alpha+\myfrac12\tv\,\xi_\alpha=0\,.
\end{equation}

For convenience, we list here the explicit forms of the 
covariant derivatives appearing in these transformation laws:
\begin{eqnarray}
\calD_\mu\varepsilon^i&=&\left(\partial_\mu-\myfrac14\gamma_{ab}\omega^{ab}_\mu\right)\varepsilon^i
-\tilde V_\mu{}^i{}_j\varepsilon^j, \nn
\hatD_\mu\tilde t^{ij}&=&  \calD_\mu\tilde t^{ij}
-4i\bar\psi_\mu^{(i}\tilde\chi^{j)}-i\bar\psi_\mu^{(i}\gamma\dt  \hatR{}^{j)}(Q)
-\myfrac{2i\calN_{IJ}}{3\calN}\left(\bar\psi_\mu^{(i} \slashD M^I\lambda^{j)J}
-\bar\psi_\mu\lambda^I\ty^{Jij}\right), \nn
\hatD_\mu\tilde v_{ab}&=& \calD_\mu\tilde v_{ab}
+2i\bar\psi_\mu\gamma_{ab}\tilde\chi 
+\myfrac{i}2\bar\psi_\mu\gamma_{ab}\gamma\dt \hatR(Q)
-\myfrac{i}4\bar\psi_\mu\gamma_{abcd} \hatR{}^{cd}(Q)
+i\bar\psi_\mu\hatR_{ab}(Q)\,, \nn
\calD_\mu\tilde t^{ij}&=&
\partial_\mu\tilde t^{ij}-2\tilde V_\mu^{(i}{}_k\tilde t^{j)k}, 
\qquad \qquad 
\calD_\mu\tilde v_{ab}=\partial_\mu\tilde v_{ab}
+2\omega_\mu{}_{[a}{}^c\tilde v_{b]c}\,, \nn
\hatD_\mu M^I&=& \calD_\mu M^I-2i\bar\psi_\mu\lambda^I, 
\qquad \qquad
\calD_\mu M^I=\partial_\mu M^I-g[W_\mu,\,M]^I, \nn 
&&\hspace{-4em}\hatD_\mu{\hat F}_{ab}(W)^I= \calD_\mu{\hat F}_{ab}(W)^I
-4i\bar\psi_\mu\gamma_{[a} \hatD_{b]}\lambda^I-2i\bar\psi_\mu\hatR_{ab}(Q)M^I 
-4i\bar\psi_\mu\gamma\dt \tilde v\gamma_{ab}\lambda^I \nn
 &&- \ 8i\bar\psi_\mu\lambda^I\tilde v_{ab} 
-4i\bar\psi_\mu\gamma_{ab}\tilde t\lambda^I 
-\myfrac{\calN_I}{3\calN}i\bar\psi_\mu\gamma\dt  {\hat
F}(W)^I\gamma_{ab}\lambda^I 
+\myfrac{8\calN_{IJ}}{3\calN}(\bar\psi_\mu\lambda^J)\bar\lambda^I\gamma_{ab}
\lambda^K, \nn
&&\hspace{-4em}\calD_\mu{\hat F}_{ab}(W)^I=\partial_\mu{\hat F}_{ab}(W)^I
-g[W_\mu,\,{\hat F}_{ab}(W)]^I+2 \omega_\mu{}_{[a}{}^c {\hat F}_{b]c}(W)^I, \nn
\hatD_\mu\lambda^I_i&=&\calD_\mu\lambda^I_i
+\calP^I{}_J(\myfrac14\gamma\dt {\hat F}(W)^J\psi_{\mu i}
+\myfrac12\slashD M^J\psi_{\mu i}-\ty^J_{ij}\psi_\mu^j) 
+\myfrac{M^I\calN_{JK}}{3\calN}(2i\bar\psi_\mu\lambda^J)\lambda^K_i, \nn
\calD_\mu\lambda^I_i&=&\left(\partial_\mu-\myfrac14\gamma_{ab}\omega^{ab}_\mu\right)\lambda^I_i
-\tilde V_{\mu ij}\lambda^{Ij}-g[W_\mu,\,\lambda_i]^I, \nn
\hatD_\mu\calA^i_\alpha&=&\calD_\mu\calA^i_\alpha-2i\bar\psi_\mu^i\xi_\alpha\,,
\qquad \qquad \calD_\mu\calA^i_\alpha=\partial_\mu\calA^i_\alpha -\tilde
V_\mu^i{}_j\calA^j_\alpha-gW_{\mu\alpha\beta}\calA^{\beta i}, \nn
\hatD_\mu\xi_\alpha&=&\calD_\mu\xi_\alpha +{\slashD}\calA_\alpha^i\psi_{\mu
i}-\psi_{\mu i}gM\calA^i_\alpha-{\mbf\Gamma}'\psi_{\mu i}\calA^i_\alpha
-\left(1+\myfrac{\slash A}\alpha\right)\psi_{\mu i}\calF^i_\alpha\,,\nn
\calD_\mu\xi_\alpha&=&\left(\partial_\mu-\myfrac14\gamma_{ab}\omega^{ab}_\mu\right)\xi_\alpha
-gW_{\mu\alpha\beta}\xi^\beta. \end{eqnarray}

The supercovariant curvatures $\hatR_{\mu\nu}$ are obtained from 
$[\hatD_a,\,\hatD_b] 
= -\hatR_{ab}{}^{\bar A}\XX_{\bar A}$ as noted above, or 
can be read directly from the above transformation laws of the gauge field, 
(\ref{eq:gaugeTrf}), via the formulas (I\,2$\cdot$29), 
$\hatR_{\mu\nu}{}^{\bar A}
=2\partial_{[\mu}h_{\nu]}^{\bar A}
-h_\mu^{\bar C}h_\nu^{\bar B}f'_{\bar B \bar C}{}^{\bar A}$, and 
(I\,2$\cdot$24), $\delta h_\mu^{\bar A} = \partial_\mu\varepsilon^{\bar A}
+\varepsilon^{\bar C}h_\mu^{\bar B}f_{{\bar B}{\bar C}}{}^{\bar A}$.
Explicitly, they are given by
\begin{eqnarray}
 \hatR_{\mu\nu}{}^i(Q)&=&2 {\calD}_{[\mu}\psi_{\nu]}^i
+2\gamma_{[\mu}\tilde t^i{}_j\psi_{\nu]}^j
+\gamma_{[\mu ab}\psi_{\nu]}^i\tilde v^{ab} \nn
&&
{}+\myfrac{\calN_I}{6\calN}\gamma_{[\mu}\gamma\dt {\hat F}^I(W)\psi_{\nu]}^i
+\myfrac{4i\calN_{IJ}}{3\calN}\gamma_{[\mu}\lambda^{Ii}(\bar\lambda^J\psi_{\nu]})\,,\nn
\hatR_{\mu\nu}{}^{ij}(U)&=&2\partial_{[\mu}\tilde V_{\nu]}^{ij}
-[\tilde V_\mu,\,\tilde V_\nu]^{ij}
+8i \bar\psi_{[\mu}^{(i}\gamma_{\nu]}\tilde\chi^{j)}
+2i \bar\psi_{[\mu}^{(i}\gamma_{\nu]ab} \hatR{}^{abj)}(Q)\nn
&&-4i\bar\psi_{\mu}^{(i}\gamma\dt \left(\tilde v
 +\myfrac{\calN_I}{4\calN} {\hat F}^I(W)\right) \psi{}_{\nu}^{j)}
+6i \bar\psi_{\mu}{ \psi}_{\nu}\tilde{t}^{ij}+\myfrac{8\calN_{IJ}}{\calN}
 (\bar\psi^\ispan_{[\mu}\lambda^I) \bar\psi_{\nu]}^{(i}\lambda^{j)J},\nn
{\hat F}^I_{\mu\nu}(W)&=&F_{\mu\nu}^I(W)
+4i \bar\psi_{[\mu}\gamma_{\nu]}\lambda^I-2i \bar\psi_\mu\psi_\nu M^I.
\end{eqnarray}

\vspace*{3mm}

\section{Compensators, gauged supergravity and scalar potential}

\subsection{Independent variables}

We have labeled the vector multiplet $(M^I, W_\mu^I, \lambda^{Ii}, \ty^{Iij})$ 
by the index $I$, taking $1+n$ values 
from 0 to $n$. However, it is only the vector component $W_\mu^I$ 
that actually has $1+n$ independent components. All the others have 
only $n$ components, since the scalar components $M^I$ satisfy 
the $\XD$ gauge condition $\calN(M)=1$, and the fermion and auxiliary 
fields satisfy the constraints $\calN_I\lambda^I=\calN_I\ty^I=0$. Thus 
our parametrizations for them are redundant, although the gauge symmetry 
is realized linearly for these variables, and hence is more manifest there. 

It is, of course, possible to parametrize these fields with independent 
variables, as was done by GST from the beginning in their on-shell 
formulation.\cite{ref:GST} 
GST parametrized the manifold $\calM$ of the scalar 
fields by $\phi^x$ with curved index $x=1,\cdots,n$, and the fermions by 
$\lambda^a$ with tangent index $a=1,\cdots,n$. We can assign the same tangent 
index to our auxiliary fields and write $\ty^a$.

The basic correspondence between the GST parametrization and ours is 
as follows:
\begin{equation}
\begin{array}{ccc}
 \hbox{GST parametrization}       &       &  \hbox{our parametrization} \\
 \calN=C_{IJK}h^I(\phi)h^J(\phi)h^K(\phi)  & \leftrightarrow & 
\calN=c_{IJK}M^IM^JM^K  \nn
h^I(\phi)  & = & -\sqrt{\myfrac23}M^I|_{\calN=1} \\
h_I(\phi)  & = & -\myfrac1{\sqrt6}\calN_I|_{\calN=1}
\end{array}
\label{eq:relGST1}
\end{equation}
From this, various geometrical quantities defined by GST can be 
translated into their counterparts in our formulation. 
The metric $a_{IJ}$ of the ambient $1+n$ 
dimensional space is the same as ours, and the metric $g_{xy}$ of the 
scalar manifold $\calM$, induced from $a_{IJ}$, is given by 
\begin{equation}
g_{xy}\equiv a_{IJ}h^I_xh^J_y\,, \qquad \hbox{with} \quad 
h^I_x\equiv-\sqrt{\myfrac32}h^I_{,x}=M^I_{,x}\,,
\end{equation}
where $,x$ denotes differentiation with respect to $\phi^x$. 
The indices $I, J,\cdots$ 
are raised and lowered by the metric $a_{IJ}$ and its inverse $a^{IJ}$, 
and the indices $x,y,\cdots$ are raised and lowered by the metric 
$g_{xy}$ and its inverse $g^{xy}$. The curved indices $x,y,\cdots$ are 
converted into the tangent indices $a,b,\cdots$ by means of the vielbein
$f^a_x$ and its inverse $f_a^x$, satisfying 
$f^a_xf^b_y\delta_{ab}=g_{xy}$ and $f^a_xf^b_yg^{xy}=\delta^{ab}$. Some 
useful relations are
\begin{eqnarray}
&&h_{Ix}\equiv a_{IJ}h^I_x=\sqrt{\myfrac32}h_{I,x}\,,
\qquad h^I_a\equiv f_a^xh^I_x\,,\qquad
T_{xyz}\equiv C_{IJK}h^I_xh^J_yh^K_z, \nn
&&h_Ih^I=1,\qquad h_I^xh^I_y=\delta^x_y,\qquad h_Ih^I_x=h^Ih_I^x= 0, \nn
&&a^{IJ}= g^{xy}h^I_xh^J_y +h^Ih^J \quad \rightarrow\quad 
\delta^I_J= g^{xy}h^I_xh_{Jy} +h^Ih_J=
\calP^I_J+{M^I\calN_J\over3\calN}\,,  \nn
&&a_{IJ}h^I_ah^J_{b,x}= \Omega_{xab}-\sqrt{\myfrac23}T_{abx}\,, \qquad 
\half a_{IJ,x}h^I_ah^J_b=\sqrt{\myfrac23}T_{abc}f^c_x\,, 
\label{eq:relGST2}
\end{eqnarray}
where $\Omega_x^{ab}$ is the `spin-connection' of $\calM$ defined as usual by 
$f^a_{[x,y]}+\Omega_{[y}^{\ ab}f^b_{x]}=0$.

Now it is easy to rewrite our action and supersymmetry transformation laws
in terms of the independent variables $\phi^x, \lambda_i^a$ and $\ty_{ij}^a$. 
$M^I$ is 
simply $-\sqrt{2/3}h^I(\phi)$, and the indices $I$ and $a$ of $\lambda$ and $\ty$ 
are mutually converted by 
\begin{equation}
\lambda^a = h^a_I\lambda^I, \qquad 
\lambda^I=\calP^I_J\lambda^J =h^I_ah_{J}^a\lambda^J=h^I_a\lambda^a.
\end{equation}
For instance, the supersymmetry transformation laws (\ref{eq:vecSuper}) 
are rewritten as
\begin{eqnarray}
\delta W^I_\mu&=&-2ih^I_a\bar\varepsilon\gamma_\mu\lambda^a-i\sqrt6h^I\bar\varepsilon\psi_\mu\,,\nn
\delta\phi^x&=&2if^x_a\bar\varepsilon\lambda^a,\nn
\delta\lambda^a_i&=&-\myfrac14h^a_I\gamma\dt {\hat F}^I(W)\varepsilon_i
-\myfrac12f^a_x\slashD\phi^x\varepsilon_i+\tilde Y^a_{ij}\varepsilon^j
-(\Omega_x^{ab}-\sqrt{\myfrac23}T_x{}^{ab})\delta\phi^x\lambda_{bi}\,,\nn
\delta\tilde Y^a_{ij}&=&
2i\bar\varepsilon_{(i}\slashD\lambda^b_{j)}
+2i\bar\varepsilon_{(i}\left(\sqrt{\myfrac32}h^IgL_I{}^a{}_b
+(\Omega_x{}^a{}_b-\sqrt{\myfrac23}T_x{}^a{}_b)\slashD\phi^x\right)\lambda^b_{j)} \nn
&&{}
+i\myfrac1{\sqrt6}h_I\bar\varepsilon_{(i}\gamma\dt {\hat F}^I(W)\lambda^a_{j)}
-i\bar\varepsilon_{(i}\gamma\dt \tv\lambda^a_{j)}
-2i\bar\varepsilon_{(i}\tilde t_{j)}{}^k\lambda^a_k+4i(\bar\varepsilon\lambda^a)\tilde t_{ij}
\nn
&&{}-\myfrac83\bar\varepsilon\lambda_b(\bar\lambda^a_{(i}\lambda^b_{j)})
-(\Omega_x{}^a{}_b-\sqrt{\myfrac23}T_x{}^a{}_b)\delta\phi^x \ty^b_{ij}\,.
\label{eq:GSTSuper}
\end{eqnarray}
Here, $L_I{}^a{}_b(\phi)$ is a function of $\phi^x$ appearing in the 
gauge transformation in the GST notation:
\begin{eqnarray}
&&\delta_G(\theta)\phi^x= gK_I^x(\phi)\theta^I, \qquad 
\delta_G(\theta)\lambda^a= gL_I{}^a{}_b(\phi)\lambda^b \theta^I, \nn
&&K_I^x(\phi)=-\sqrt{\myfrac32}h^x_Kf_{JI}{}^Kh^J, \nn
&&L_I{}^a{}_b(\phi)=
-(\Omega_x{}^a{}_b-\sqrt{\myfrac23}T_x{}^a{}_b)K^x_I+h_K^af_{JI}{}^Kh^J_b.
\end{eqnarray}

One can see that these transformation laws for the physical components 
$W_\mu^I, \phi^x$ and $\lambda_i^a$ agree with the GST result\cite{ref:GST} 
if the auxiliary fields 
are replaced by their solutions [and $2\lambda^a$, $2\psi_\mu$, $2\varepsilon$ and 
$i\gamma_\mu(-i\gamma^\mu)$ here 
are identified with $\lambda^a$, $\psi_\mu$, $\varepsilon$ and $\gamma_\mu(\gamma^\mu)$ of GST.] \
One can also easily rewrite the action and see the agreement with GST 
for the on-shell part in the absence of the hypermultiplet. 

In the case of the hypermultiplet, $\calA_\alpha^i$ and $\xi_\alpha$ are
independent variables off-shell. However, on-shell they  become 
mutually dependent variables, since they satisfy the equations of motion
$\calA^2=-2$ and $\calA^{\bar\alpha}_i\xi_\alpha=0$. Moreover, there 
remains the $SU(2)$ $\XU$ gauge symmetry, with which three components of 
$\calA_\alpha^i$ can be eliminated. (Thus at least four of the 
$\calA_\alpha^i$ 
and two of the $\xi_\alpha$ can be eliminated. Generally, {\it 
compensator} components of the hypermultiplets can be eliminated by 
equations of motion and the gauge symmetries, as explained below.) It is
possible to separate the variables even off-shell into 
genuine independent variables and other variables that 
vanish on-shell or can be eliminated by gauge fixing. Such independent 
variables are those used in the on-shell formulation, for instance, by 
Ceresole and Dall'Agata,\cite{ref:CDA} and they are formally very 
similar to the GST variables for vector multiplets. Hence, the rewriting
of the hypermultiplet variables can be done in a manner similar to that 
in the vector multiplet case. The only complications in this case are 
the above mentioned separation of the on-shell (or gauge) vanishing variables, 
which depend on the number of the compensators (i.e., the structure of 
the hypermultiplet manifold).

\subsection{Compensator}

The $\XD$ gauge fixing $\calN=1$ was necessary to obtain the canonical 
form of the Einstein-Hilbert term. Owing to the equation of motion 
$\calA^2+2\calN=0$, this in turn implies that the relation
\begin{equation}
\calA^2\equiv\calA_i^\alpha d_\alpha{}^\beta\calA_\beta^i=-2
\label{eq:Dconstr}
\end{equation}
must hold 
on-shell. But this is possible only if some components of the 
hypermultiplet $\calA^\alpha_i$ have {\it negative metric}.\cite{ref:dWvHVP} 
To see this, we recall the fact that the metric $d_\alpha{}^\beta$ of 
the hypermultiplet can be brought into the standard form\cite{ref:dWLVP}
\begin{equation}
d_\alpha{}^\beta=\pmatrix{{\bf{1}}_{2p}&  \cr
        &-{\bf{1}}_{2q} \cr}.
\qquad (p,q: \hbox{integer})
\end{equation}
We distinguish the first $2p$ components of the hypermultiplet 
$\calA^\alpha_i$ with index $\alpha=1,2,\cdots,2p$ from the rest of the 
$2q$ components, and use the indices $a$ and $\6\alpha$ to denote the 
former $2p$ and the latter $2q$ components, respectively. Also taking 
account of the hermiticity $\calA^i_\alpha=-(\calA^\alpha_i)^*$, the 
quadratic terms of the hypermultiplet read
\begin{eqnarray}
&&\calA^2\equiv\calA_i^\alpha d_\alpha{}^\beta\calA_\beta^i
=-(\calA^a_i)^*(\calA^a_i)+(\calA^\6\alpha_i)^*(\calA^\6\alpha_i) 
\equiv-|\calA^a_i|^2+ |\calA^\6\alpha_i|^2 ,\nn
&&\nabla^\mu\calA^{\bar\alpha}_i\nabla_\mu\calA_\alpha^i 
=-(\nabla^\mu\calA^{a}_i)^*(\nabla_\mu\calA^a_i) 
+(\nabla^\mu\calA^{\6\alpha}_i)^*(\nabla_\mu\calA^\6\alpha_i). 
\end{eqnarray}
Thus we see that the first $2p$ components $\calA^a_i$ (corresponding to 
$p$ quaternions) have negative metric and hence should not be physical 
fields. Indeed, they are so-called {\it compensator} fields, which are 
used to fix the extraneous gauge degrees of freedom. In the simplest case, 
$p=1$, for instance, the compensator $\calA^a_i$ has four real 
components, among which one component is already eliminated by the above
condition (\ref{eq:Dconstr}). The remaining three degrees of freedom can
also be eliminated by fixing the $SU(2)$ $\XU$ gauge by the condition
\begin{equation}
\calA^a_i \propto\delta^a_i\quad \rightarrow\quad 
\calA^a_i = \delta^a_i \sqrt{1+\half|\calA^\6\alpha_i|^2} =-\calA^i_a.
\label{eq:comp}
\end{equation}
The target manifold $\calM_Q$ of the scalar fields $\calA^\alpha_i$ becomes 
$USp(2,2q)/USp(2)\times USp(2q)$ in this case. 
For $p\geq2$, we need to have more gauge freedom to eliminate more 
negative metric fields. In particular, if we add vector multiplets 
which couple to the hypermultiplet but do not have their own kinetic terms, 
the corresponding auxiliary fields $Y^{ij}$ do not have quadratic terms 
and act as multiplier fields to impose further constraints on the 
scalar fields $\calA^\alpha_i$ on-shell.\footnote{%
The corresponding fermion component, the gaugino $\lambda^i$, also becomes a
multiplier to impose a constraint on the hypermultiplet fermion fields 
$\xi_\alpha$.} 
For instance, it is known that 
the manifold $SU(2,q)/SU(2)\times SU(q)\times U(1)$ is realized for 
$p=2$ by adding a $U(1)$ vector multiplet without a kinetic 
term.\cite{ref:BS} (See Appendix B for a detailed explanation.)  \ This 
manifold reduces to $SU(2,1)/SU(2)\times U(1)$ when $q=1$, which is the 
manifold for the universal hypermultiplet appearing in the reduction of 
the heterotic M-theory on $S^1/Z_2$ to five dimensions.\cite{ref:Lukas}

\subsection{$SU(2)_R$ or $U(1)_R$ gauging}

The so-called gauged supergravity is the supergravity in which the 
$R$ symmetry $G_R$ is gauged, and $G_R$ may be either the $U(1)$ 
subgroup\cite{ref:AFMR} 
or the entire $SU(2)$ group,\cite{ref:NSdWVP} 
which act on the indices $i$ of $\psi^i_\mu$, $\lambda^{Ii}$ 
and $\calA^\alpha_i$. In our framework, this $SU(2)$ is 
already the gauge symmetry $\XU$, whose gauge field is $V_\mu^{ij}$. 
However, this gauge field $V_\mu^{ij}$ has no kinetic term and is an 
auxiliary field. To obtain a physical gauge field possessing a kinetic 
term, we must prepare another gauge field $W_\mu^R{}^a{}_b$ for $G_R$,
under which only the compensator field $\calA^a_i$ is charged:
\begin{equation}
\calD_\mu\calA^a_i = \partial_\mu\calA^a_i - V_{\mu ij}\calA^{aj}
-g_RW_{R\mu}{}^a{}_b \calA^b_i. 
\label{eq:covder}
\end{equation}
In this expression, we are assuming that the compensator has no group 
charges other than $G_R$ and that $p=1$  so that the index $a$ runs over 1 
and 2. The generator $t_R$ of $G_R$ is given by $i\vec{\sigma}^a{}_b$ in
the case of $SU(2)_R$ with the Pauli matrix $\vec{\sigma}$, and by 
$i\vec{q}\cdot\vec{\sigma}^a{}_b$ with an arbitrary real 3-vector 
$\vec{q}$ of unit length $|{\vec{q}}\,|=1$ in the case of $U(1)_R$:
\begin{equation}
W_{R\mu}{}^a{}_b = \cases{
\vec{W}_{R\mu}\cdot i\vec{\sigma}^a{}_b &for $SU(2)_R$, \cr
{W}_{R\mu}\,i\vec{q}\cdot\vec{\sigma}^a{}_b &for $U(1)_R$. \cr}
\end{equation}

It should be noted that the $G_R$ gauging interferes with the 
possibility of a hypermultiplet mass term. Indeed, the symmetric tensor 
$\eta^{\alpha\beta}$ of the mass term (\ref{eq:hypermass}) must be 
invariant under $G$, implying the constraint $[t_I,\eta]=0$ on the 
matrix $\eta=(\eta^\alpha{}_\beta) $ for any generators $t_I$ of $G$. In
particular, for the generator $t_R$ of $G_R$, which we are now assuming 
to rotate only the compensator components $\calA^a_i$, this constraint 
implies that the $2\times2$ matrix $
\eta^a{}_b$ in the compensator sector must commute with the above $t_R$. 
However, for the $G_R=SU(2)_R$ case, there is no such $\eta^a{}_b$ that 
commutes with all the Pauli matrices, 
so that the mass term cannot exist for the 
compensator. For the $G_R=U(1)_R$ case, on the other hand, the 
constraint allows $\eta^a{}_b\propto i\vec{q}\cdot\vec{\sigma}^a{}_b$. The mass term 
with this $\eta$ yields, in the above $\calD_\mu\calA^a_i$,
an additional `central charge term' $-A_\mu(gt_{I=0})^a{}_b\calA^b_i$, 
with $gt_0$ defined in Eq.~(\ref{eq:undstd0}). However, since 
$\eta^a{}_b\propto i\vec{q}\cdot\vec{\sigma}^a{}_b$, this term can be 
absorbed into the 
$-g_RW_{R\mu}^a{}_b\calA^b_i$ term, and Eq.~(\ref{eq:covder}) remains 
unchanged.  
Generally speaking, the $U(1)_R$-gauge field $W_{R\mu}$ is, of course, 
a member of our 
complete set of vectors $\{\, W^I_\mu\,\}$ and is given by a
linear combination of the latter as
\begin{equation}
W_{R\mu}= V_I W_\mu^I,
\end{equation}
with real coefficients $V_I$, which are non-vanishing only for the 
Abelian indices $I$. Therefore, if the mass term exists with 
$\eta^a{}_b=i\vec{q}\cdot\vec{\sigma}^a{}_b$, it is implied that 
the $I=0$ coefficient $V_0$ is given by $g_RV_{I=0}=m/2$.


The gauge fields $V_\mu$ and $W_{R\mu}$ mix with each other. 
We redefine the $\XU$ gauge field $V_\mu{}^i_j$ as
\begin{equation}
V_\mu^{{\rm N}ij} \equiv 
V_\mu^{ij} -g_RW_{R\mu}{}^{ij},
\end{equation}
while keeping the $SU(2)_R$ gauge field $W_{R\mu}$ intact. Then, 
noting the $SU(2)$ $\XU$ gauge-fixing condition $\calA^a_i\propto\delta^a_i$, 
we see that the compensator couples only to this new 
$SU(2)$ gauge field $V_\mu^{\rm N}$ and no longer couples to 
the $SU(2)_R$ gauge field $W_{R\mu}$:
\begin{equation}
\calD_\mu\calA^a_i = (\delta^a_i\partial_\mu+ V_\mu^{\rm N}{}^a_i)
\sqrt{1+\half|\calA^\6\alpha_i|^2}.
\end{equation}
On the other hand, other fields carrying the original $SU(2)$ indices $i$ 
now come to couple both to $V_\mu^{\rm N}$ and $W_{R\mu}$, 
since $V_\mu$ should now be replaced by $V_\mu^{\rm N}+g_RW_{R\mu}$. 
Therefore the net effect of the $SU(2)_R$ [or $U(1)_R$] gauging is 
simply that 1) the auxiliary field $V_\mu$ is replaced by $V_\mu^{\rm 
N}$, and 2) the covariant derivative $\nabla_\mu$ (or $\calD_\mu$) should
be understood to contain the $W_{R\mu}$ covariantization term 
$-\delta_R(W_{R\mu})$ if acting on the fields carrying the $SU(2)$ 
indices $i$. The previously derived action remains valid as it stands 
with this understanding. 

\subsection{Scalar potential}

The scalar potential term can be read from the action 
(\ref{eq:finalAction}) to be
\begin{equation}
V = 
\myfrac14(a^{IJ}-M^IM^J)
\calY_I^{ij}\calY_{Jij}|_{\rm bosonic\ part}
-\calA^{\bar\alpha}_i(gM)^2{}_\alpha{}^\beta\calA_\beta^i.
\end{equation}
Here the first term has come from the elimination of the auxiliary 
fields $Y^{Iij}$ of the vector multiplet and $t^{ij}$ of the Weyl 
multiplet, and the second term from the hypermultiplet. Using 
Eq.~(\ref{eq:sol}) for $\calY_I^{ij}$, this potential can be rewritten 
in the form
\begin{eqnarray}
V &=&(a^{IJ}-M^IM^J)P_I^{ij}P_{Jij} +Q^{\bar\alpha}_iQ^i_\alpha\nn
&=&(a^{IJ}-M^IM^J)P_I^{ij}(P_J^{ij})^* -|Q^{a}_i|^2+|Q^{\6\alpha}_i|^2\,,
\label{eq:SPotential}
\end{eqnarray}
where
\begin{equation}
P_I^{ij}\equiv\calA^{(i}_\alpha gt_I^{\bar\alpha\beta}\calA^{j)}_\beta 
=d_\gamma{}^\alpha\calA^{(i}_\alpha gt_I^{\gamma\beta}\calA^{j)}_\beta, \qquad 
Q^\alpha_i \equiv g\delta_G(M)\calA^\alpha_i = 
M^I(gt_I)^\alpha{}_\beta\calA^\beta_i,
\end{equation}
%
and we have used the hermiticity properties 
$(P_I^{ij})^* = P_{Iij}$ and $Q^i_\alpha=-(Q^\alpha_i)^*$.
Since $a_{IJ}$ is the metric of the vector multiplet, 
the first term $a^{IJ}P_I^{ij}(P_J^{ij})^*$ is positive definite. 
Negative contributions result from the terms $-|M^IP_I^{ij}|^2$ and 
$-|Q^{a}_i|^2$, the latter of which comes from the compensator component 
of the hypermultiplet. 

Equation (\ref{eq:SPotential}) is our general result for the scalar 
potential. Consider here the special case of $U(1)_R$-gauged 
supergravity in which $p=1$ and $q=0$; that is, there is a single 
(quaternion) compensator and no physical hypermultiplets. Then, the 
compensator $\calA^a_i$ becomes simply a constant $\delta^a_i$, by 
Eq.~(\ref{eq:comp}). If the compensator is charged only under the 
$U(1)_R$ in $G$, we have 
\begin{eqnarray}
P_I^{ij}&=&\calA^{(i}_agt_I^{ab}\calA^{j)}_b
=g_RV_I\epsilon^{jk}(i\vec{q}\cdot\vec{\sigma})^i{}_k 
\,, \nn
Q^a_i &=& M^IV_Ig_R(i\vec{q}\cdot\vec{\sigma})^a{}_i 
\,,
\end{eqnarray}
and the scalar potential
\begin{eqnarray}
V&=&2g_R^2(a^{IJ}-2M^IM^J)V_IV_J 
=2g_R^2(g^{xy}h^I_xh^J_y-2h^Ih^J)V_IV_J  \nn
&=&g_R^2\left({9\over2}g^{xy}{\partial W\over\partial\varphi^x}{\partial W\over\partial\varphi^y}
-6W^2\right),
\end{eqnarray}
where we have used the relations $a^{IJ}=g^{xy}h^I_xh^J_y+h^Ih^J$ and 
$h^I=-\sqrt{2/3}M^I$ in Eqs.~(\ref{eq:relGST1}) and (\ref{eq:relGST2}), 
and the definitions\cite{ref:BKVP}
\begin{equation}
W\equiv\sqrt{\myfrac23}h^IV_I = -\myfrac23M^IV_I\,, \qquad 
{\partial W\over\partial\varphi^x}
= -\myfrac23h^I_x V_I
= -\myfrac23M^I_{,x} V_I\,.
\end{equation}
This agrees with the result by GST.\cite{ref:GST} 
If the physical vector multiplets 
are not contained in the system, the scalars $\varphi^x$ do not appear
either, and only the graviphoton with $I=0$ exists. In this case 
$\calN=c_{000}\alpha^3$, and $\alpha=M^{I=0}$ is determined to be $\sqrt{3/2}$ 
by the normalization requirement of the graviphoton kinetic term,
$a_{00}=1$. Then, $W=-\sqrt{2/3}V_0$,
 and hence the potential further reduces to
\begin{equation}
V=-4g_R^2V_0^2\,,
\end{equation}
which agrees with the well-known anti-de Sitter cosmological term in the 
pure gauged supergravity.\cite{ref:AFMR}

\section{Conclusion and discussion}

In this paper, we have presented an action for a general system of 
Yang-Mills vector multiplets and hypermultiplet matter fields coupled to 
supergravity in five dimensions. The supersymmetry transformation rules 
were also found. We have given these completely 
in the off-shell formulation, in which all the auxiliary fields are 
retained. Our work can be considered an off-shell extension of the 
preceding work by GST\cite{ref:GST} and its generalization by 
Ceresole and Dall'Agata.\cite{ref:CDA}  [The latter authors also included 
`tensor multiplet matter fields' (linear multiplets, in our terminology) 
with regard to which our system is less general.] 

We have several applications in mind, such as compactifying on the orbifold 
$S^1/Z_2$ and/or adding D-branes to the system. Then, the power of
the present off-shell formulation will become apparent. In particular, 
for the case of $S^1/Z_2$, it should be straightforward to determine 
how to couple the bulk fields to the fields on the boundary planes, 
since we can follow the general algorithm given by Mirabelli and Peskin for 
the case of the bulk Yang-Mills supermultiplet.\cite{ref:MP} 
Indeed, this program has been started very recently 
by Zucker\cite{ref:Zucker2} using his off-shell formulation. 
He used a `tensor multiplet' (linear multiplet) as a compensator for the 
five-dimensional (pure) supergravity and found that the 4D supergravity 
induced on the boundaries is a non-minimal version of $N=1$ Poncar\'e 
supergravity with $16+16$ components containing one auxiliary spinor, 
which was presented by Sohnius and West long ago.\cite{ref:SW} 
This non-minimal version is related to the new minimal version by the 
same authors.\cite{ref:SW2} Another version of $N=1$ Poncar\'e 
supergravity, which is related to the usual minimal 
version,\cite{ref:oldminimal}\ will appear if we start with our 5D 
supergravity in which the compensator is a hypermultiplet.

Adding D-branes in the system is not so straightforward. First of all, 
a D-brane is a dynamical object whose position $X^\mu(x)$ in the bulk and 
its fermionic counterpart become a supermultiplet in 4D that realizes 
the bulk (local) supersymmetry {non-linearly}. The problem of identifying 
a supersymmetry transformation law for this multiplet and writing an 
invariant action is already quite non-trivial, even in the case of rigid 
supersymmetry, and has long been studied by several 
authors.\cite{ref:partial} Once this problem is settled, coupling the 
bulk supergravity to the fields on the D-brane should be easy also in this 
case. The off-shell formulation is essential in any case.

\section*{Acknowledgements}
The authors would like to thank Andr\'e Lukas, Hiroaki Nakano, 
Paul Townsend, Antoine Van~Proeyen, Bernard de~Wit and Max Zucker for 
discussions and useful information. They also appreciate the Summer 
Institute 2000 held at Fuji-Yoshida, at which a preliminary version of 
this work was reported. T.~K.\ is supported in part by a Grant-in-Aid 
for Scientific Research (No.~10640261) from the Japan Society for the 
Promotion of Science and a Grant-in-Aid for Scientific Research on 
Priority Areas (No.~12047214) from the Ministry of Education, Science, 
Sports and Culture, Japan.

\newpage
\appendix

\section{A Representation Realizing 
Eq.~(\protect\ref{eq:CIJK})}

The following is an example of the set of hermitian matrices $\{ T_I \}$, 
realizing the property (\ref{eq:CIJK}).

Let us prepare  a representation vector $\psi_i$ 
for each simple factor group $G_i$ in $G$ that gives a faithful 
representation $R_i$ of $G_i$, and a suitable numbers of singlet vectors
$\{ \psi_\alpha\}$. Assigning to them suitable $U(1)_x$ charges also,
we consider a representation of $G$ whose representation vector is given by 
$\{ \psi_j,\ \psi_\alpha\}$, which transforms as follows under 
$G=\prod_{i}G_i\times\prod_{x}U(1)_x$:
\vspace{1ex}
\begin{center}
\begin{tabular}{c|c|c} \hline 
        &   under $G_i$      &   $U(1)_x$ charges    \\ \hline
$\psi_j$  &   repr.\ $R_j$ for $i=j$ and singlet for $i\not=j$  & $q^x_j$ \\ \hline
$\psi_\alpha$  &   singlet    &  $q^x_\alpha$     \\ \hline
\end{tabular}
\vskip2ex
\end{center}
Let $A_i$ be the generator label of the simple factor group $G_i$, $a_i$ be 
the component label of the $\dim R_j$ vector $\psi_j=(\psi_j^{a_i})$, and 
$\rho_{R_i}(t_{A_i})=({\rho_{R_i}(t_{A_i})}^{a_i}{}_{b_i})$ 
be the representation matrices of the generators acting on 
$\psi_i$ in the representation $R_i$. Then the generators 
$t_I=(t_{A_i}, t_x)$ of $G$ are 
given in this representation by 
\begin{eqnarray}
&&t_{A_i}{}^{a_j}{}_{b_j}=\delta_{ij}{\rho_{R_i}(t_{A_i})}^{a_i}{}_{b_i}, \qquad 
t_{A_i}{}^{\alpha}{}_{\beta}=0, \nn
&&t_{x}{}^{a_j}{}_{b_j}=i\delta^{a_j}_{b_j}q^x_j, \qquad \qquad \qquad 
t_{x}{}^{\alpha}{}_{\beta}=i\delta^\alpha_\beta q^x_\alpha. 
\end{eqnarray}
The desired matrices $T_I$ are given by $T_{A_i}=c_it_{A_i}/i$ and 
$T_x=t_x/i$.
Equations given by (\ref{eq:CIJK}) to be satisfied are
\begin{eqnarray}
G_i^3 &:& \quad 6c_{A_iB_iC_i}= 
-ic_i^3 \tr\bigl(\rho_{R_i}(t_{A_i})
\{\rho_{R_i}(t_{B_i}),\rho_{R_i}(t_{C_i})\}\bigr), \nn
G_i^2U(1)_x&:& \quad 3c_{A_iB_ix}= 
-c_i^2 q^x_i \tr\bigl(\rho_{R_i}(t_{A_i})\rho_{R_i}(t_{B_i})\bigr), \nn
U(1)_xU(1)_yU(1)_z&:& \quad 3c_{xyz}= 
 \sum_iq^x_iq^y_iq^z_i\dim R_i + \sum_\alpha q^x_\alpha q^y_\alpha q^z_\alpha. \nonumber
\end{eqnarray}
The constants $c_i$ and $U(1)_x$ charges $q^x_i$ of $\psi_i$ are fixed by the 
first and second equations, respectively. The third equation should be 
satisfied by adjusting the 
$U(1)_x$ charges $q^x_\alpha$ of $\psi_\alpha$. Clearly, there are such 
solutions for $q^x_\alpha$ if there are sufficiently many $\psi_\alpha$.

\section{$U(2,n)/U(2)\times U(n)$ as a Hypermultiplet Manifold for $p=2$}

In this appendix we explain how the 
manifold $U(2,n)/U(2)\times U(n)$ appears as a target space manifold 
$\calM_Q$ of the physical hypermultiplet scalar fields for the case 
$p=2$. This is merely a detailed version of what was essentially 
shown long ago by Breitenlohner and Sohnius.\cite{ref:BS}

We consider the hypermultiplet $\calA^\alpha_i$ in the standard 
representation, in which the matrices $d_\alpha{}^
\beta$ and $\rho^{\alpha\beta}$ take the form\cite{ref:dWLVP}
\begin{equation}
d_\alpha{}^\beta=\pmatrix{{\bf{1}}_{2p}&  \cr
        &-{\bf{1}}_{2q} \cr}, \qquad 
\rho^{\alpha\beta}=\rho_{\alpha\beta}=\pmatrix{\epsilon&&  \cr
        &\epsilon& \cr
        && \ddots}. \quad (\epsilon\equiv i\sigma_2) 
\end{equation}
The hypermultiplet $\calA_{\alpha i}$ is regarded as the 
$2(p+q)\times2$ matrix 
\begin{equation}
\calA = \pmatrix{ \cr \calA_{\alpha i} \cr \cr}
= \pmatrix{\vdots \cr \noalign{\vskip.8ex}
\matrix{\calA_{2a-1,1}&\calA_{2a-1,2}\cr
        \calA_{2a,1}&\calA_{2a,2}\cr} 
 \cr \vdots}, \quad (a=1,2,\cdots,p+q)
\end{equation}
which consists of $p+q$ $2\times2$-blocks. Each block can be identified 
with a quaternion, which is also mapped equivalently to a $2\times2$ 
matrix:
\begin{equation}
{\mbf q}\equiv q^0+{\mbf i}q^1+{\mbf j}q^2+{\mbf k}q^3
\quad \leftrightarrow \quad 
q^0{\mbf1}_2-i\vec{q}\cdot\vec{\sigma}= 
\pmatrix{q^0-iq^3 & -iq^1-q^2 \cr
 -iq^1+q^2 & q^0+iq^3 \cr}.
\end{equation}
This is consistent with the hermiticity condition for the 
hypermultiplet:
\begin{equation}
(\calA_{\alpha i})^*=\calA^{\alpha i}=\rho^{\alpha\beta}\epsilon^{ij}\calA_{\beta j}, \quad 
\rightarrow\quad \calA^\dagger= -\epsilon\calA^\T\rho\,. 
\end{equation}
The group $G$ transformation and $SU(2)$ $\XU$ transformation 
act on $\calA$ as
\begin{equation}
\calA \ \rightarrow\ \calA' = g\calA u^\dagger,  
\qquad g\in G, \ \ u\in SU(2).
\end{equation}
The $G$ invariance of the quadratic form 
\begin{equation}
\calA^{\alpha i}d_\alpha{}^\beta\calA_{\beta j} \quad \leftrightarrow \quad 
\calA^\dagger d\,\calA = -\epsilon\calA^\T\rho d\,\calA 
\end{equation}
requires that the two conditions for $g\in G$,
\begin{equation}
g^\dagger d\,g =d, \qquad g^\T\rho d\,g = \rho d\,,
\end{equation}
be satisfied. The former implies $g\in U(2p,2q)$ and the latter $g\in 
Sp(2p+2q;{\mbf C})$, so that the group $G$ must be a subgroup of 
$USp(2p,2q)=U(2p,2q)\cap Sp(2p+2q;{\mbf C})$. 

Now we consider the case $p=2$, in which we gauge the $U(1)$ group, 
which acts on $\calA$ as a phase rotation $e^{i\theta}$ for the odd 
rows and as $e^{-i\theta}$ for the even rows; that is, the generator 
is given by $T_3 = \sigma_3\otimes {\bf1}_{p+q}$. We do not give a kinetic 
term for the vector multiplet $\V_3$ coupling to this charge $T_3$. Then, 
the auxiliary field component $Y_3^{ij}$ of this multiplet 
appears only in a linear form in the action: 
$2Y_3{}^i_j\calA^{\alpha i}d_\alpha{}^\beta T_{3\beta}{}^\gamma\calA_{\gamma j}=
2\tr(Y_3\calA^\dagger d\,T_3\calA)$. 
Thus it acts as a multiplier to impose the following three 
constraints on the hypermultiplet on-shell: 
\begin{equation}
\tr(\sigma^a\calA^\dagger d\,T_3\calA)=0 \qquad \hbox{for} \quad a=1,2,3.
\label{eq:Yconstr}
\end{equation}
Moreover, we have one more constraint on-shell,
\begin{equation}
\tr(\calA^\dagger d\,\calA) =2\,,
\label{eq:Dconstr1}
\end{equation}
which comes from the equation of motion $\calA^2=-2\calN$ and the $\XD$ 
gauge fixing condition $\calN=1$. 
Recall that we have two quaternion compensators for the present $p=2$ case.
Hence there are eight (real) scalar fields with negative metric which should 
be eliminated. 
The above constraints eliminate four components, and we still have 
$SU(2)$ $\XU$ symmetry acting on the index $i$ and the $U(1)$ gauge 
symmetry for the charge $T_3$. 
We can eliminate the remaining four negative metric components by 
the gauge-fixing of these gauge symmetries, so that the theory is 
consistent. 

The manifold of the hypermultiplet specified by these four constraints 
(\ref{eq:Yconstr}) and (\ref{eq:Dconstr1}) have dimension 
$4(p+q)-4=4+4q$, and it is seen to be $U(2,q)/U(2)\times U(q)$ as 
follows. First, we find that a representative element of $\calA$ 
satisfying these constraints is given by
\begin{equation}
\calA^{\rm repr} = {1\over\sqrt2}\pmatrix{
{\bf1}_2 \cr
i\sigma_2 \cr
 {\bf0}_2 \cr
 \vdots \cr
 {\bf0}_2 \cr}\ .
\end{equation}
Second, to identify the manifold, it is 
sufficient to consider the half size $(p+q)\times2$ complex
 matrix $\calA_{\rm odd}$ 
that consists of the odd rows of $\calA$ alone, since  
the even row elements are essentially the complex conjugates of 
the odd row elements, as stipulated by the reality condition of $\calA$.
In this half-size representation, we can see that unitary transformations
of the above representative element,
\begin{equation}
\calA_{\rm odd} = U 
\calA_{\rm odd}^{\rm repr} = {1\over\sqrt2}U\pmatrix{
 1\ 0 \cr
 0\ 1 \cr
 0\ 0 \cr
 \vdots \cr
 0\ 0\cr}\ , \qquad 
U\in U(p,q),
\end{equation}
all satisfy the above constraints. But here, the subgroup 
$U(q)\subset U(p,q)$ rotating the lower $q$ rows alone is inactive, 
so that
the manifold of $\calA_{\rm odd}$ given by this form is $U(p,q)/U(q)$ 
and has dimension $(p+q)^2-q^2=p^2+2pq$. However, when $p=2$, this 
dimension already equals the above dimension $4q+4$ for the 
hypermultiplet $\calA$ specified by the constraints (\ref{eq:Yconstr}) 
and (\ref{eq:Dconstr1}), and thus the manifold of the latter is proved to
be $U(2,q)/U(q)$.

The manifold of the physical hypermultiplets is further reduced by 
the gauge fixing of $SU(2)$ and $U(1)$, and hence becomes 
$U(2,q)/U(q)\times U(2)$.

Note also that the gauge group $G$ is reduced to a subgroup of 
$U(p,q)$ as a result of the gauging of $U(1)$. Indeed, the 
gauge transformation $g\in G$, compatible with the $U(1)$ symmetry, should
commute with the $U(1)$ generator $T_3$: 
$gT_3=T_3g$. 
One can easily see that the group element $g$ in 
$USp(2p,2q)$ satisfying this condition further must have the form
\begin{equation}
g= \pmatrix{ U & 0 \cr 0 & U^{\T\,-1} \cr}, \quad \hbox{on}\quad 
\pmatrix{\calA_{\rm odd} \cr \calA_{\rm even}\cr}
\qquad U\in U(p,q).
\end{equation}
This element clearly belongs to $U(p,q)$.

\end{document}